\documentclass[%
 reprint,
 amsmath,amssymb,
 aps,
prstper,
floatfix,
]{revtex4-2}

\usepackage{graphicx}
\usepackage{dcolumn}
\usepackage{bm}
\usepackage{hyperref}

\begin{document}

\preprint{APS/123-QED}

\title{Thermalization rate of polaritons in strongly-coupled molecular systems}

\author{Evgeny A. Tereshchenkov}
\affiliation{ Dukhov Research Institute of Automatics (VNIIA), 22 Sushchevskaya, Moscow 127055, Russia; }
\affiliation{ Moscow Institute of Physics and Technology, 9 Institutskiy pereulok, Dolgoprudny 141700, Moscow region, Russia; }
\affiliation{ Institute for Theoretical and Applied Electromagnetics, 13 Izhorskaya, Moscow 125412, Russia; }
\author{Ivan V. Panyukov}
\affiliation{ Dukhov Research Institute of Automatics (VNIIA), 22 Sushchevskaya, Moscow 127055, Russia; }
\affiliation{ Moscow Institute of Physics and Technology, 9 Institutskiy pereulok, Dolgoprudny 141700, Moscow region, Russia; }

\author{Mikhail Misko}
\affiliation{ Moscow Institute of Physics and Technology, 9 Institutskiy pereulok, Dolgoprudny 141700, Moscow region, Russia; }

\author{Vladislav Yu. Shishkov}
\email{vladislavmipt@gmail.com}
\affiliation{ Dukhov Research Institute of Automatics (VNIIA), 22 Sushchevskaya, Moscow 127055, Russia; }
\affiliation{ Moscow Institute of Physics and Technology, 9 Institutskiy pereulok, Dolgoprudny 141700, Moscow region, Russia; }
\author{Evgeny S. Andrianov}
\affiliation{ Dukhov Research Institute of Automatics (VNIIA), 22 Sushchevskaya, Moscow 127055, Russia; }
\affiliation{ Moscow Institute of Physics and Technology, 9 Institutskiy pereulok, Dolgoprudny 141700, Moscow region, Russia; }

\author{Anton~V.~Zasedatelev}
\email{anton.zasedatelev@univie.ac.at}
\affiliation{
Vienna Center for Quantum Science and Technology~(VCQ),
~Faculty~of~Physics,~University~of~Vienna, Boltzmanngasse~5, 1090~Vienna, Austria
}

\date{\today}

\begin{abstract}
Polariton thermalization is a key process in achieving light-matter Bose--Einstein condensation, spanning from solid-state semiconductor microcavities at cryogenic temperatures to surface plasmon nanocavities with molecules at room temperature. Originated from the matter component of polariton states, the microscopic mechanisms of thermalization are closely tied to specific material properties. In this work, we investigate polariton thermalization in strongly-coupled molecular systems. We develop a microscopic theory addressing polariton thermalization through electron-phonon interactions (known as exciton-vibration coupling) with low-energy molecular vibrations. This theory presents a simple analytical method to calculate the temperature-dependent polariton thermalization rate, utilizing experimentally accessible spectral properties of bare molecules, such as the Stokes shift and temperature-dependent linewidth of photoluminescence, in conjunction with well-known parameters of optical cavities. 
Our findings demonstrate qualitative agreement with recent experimental reports of nonequilibrium polariton condensation in both ground and excited states, and explain the thermalization bottleneck effect observed at low temperatures. 
This study showcases the significance of vibrational degrees of freedom in polariton condensation and offers practical guidance for future experiments, including the selection of suitable material systems and cavity designs.

\end{abstract}

\maketitle

\section{Introduction}

Electronic and vibrational states hold a central place in the molecular systems that drive photo-induced processes in nature~\cite{jang2018delocalized} and underlie many of the technologies we interact with on a daily basis~\cite{ostroverkhova2016organic}. Despite the large energetic disparity between electronic and vibrational states, they can undergo substantial electron-phonon type interactions, known as exciton-vibration coupling~\cite{hadziioannou2006semiconducting, martinez2020dyes}. This coupling defines absorption and emission spectra of molecular systems~\cite{gierschner2003optical, kirton2015thermalization, reitz2019langevin} and shapes the relaxation dynamics at the microscopic level~\cite{bredas2004charge, coles2011vibrationally, hestand2018expanded}. When placed inside optical cavities, molecules can engage in strong light-matter interactions, leading to an effective tripartite interaction~\cite{kirton2013nonequilibrium,zasedatelev2021single, shishkov2023mapping}. This results in the formation of new eigenstates, known as vibronic-type exciton-polaritons~\cite{herrera2017dark, shishkov2023mapping}, also referred as polaron polaritons in exciton-plasmon systems~\cite{rodriguez2013thermalization, wu2016polarons}. Due to the typically strong exciton-vibration coupling, these polariton states are prevalent in molecular systems. Being hybrid light-matter states, polaritons inherit properties from both molecules and the cavity electromagnetic field. From the latter, they acquire a small effective mass and a typically short lifetime $\sim100$~fs. 

As bosons, polaritons can form Bose–Einstein condensate~(BEC), where their small effective mass and low density of states favor condensation at elevatedtemperatures~\cite{deng2010exciton}. However, their short lifetime means that polariton BEC in molecular systems is significantly out of thermal equilibrium, requiring constant re-population of the reservoir to compensate for polariton losses~\cite{imamog1996nonequilibrium}. Despite the intrinsic nonequilibrium nature of polariton condensates, they can still achieve some form of effective local equilibrium with the environment. This equilibrium is defined by the complex interplay of gain, loss, and thermalization within the polariton system. Recent theoretical reports have rigorously demonstrated this regime through an exact solution for the density matrix in the fast thermalization limit~\cite{shishkov2022exact, shishkov2022analytical}. 
The formation of polariton BEC, in general, requires two conditions: 1 - the effective rate of polariton thermalization overcomes the energy dissipation, and 2 - the total number of lower polaritons surpasses the critical number~\cite{shishkov2022exact,shishkov2022analytical}.
While the effective thermalization rate overcomes the dissipation rate for the lower polaritons above the condensation threshold, it is not always true at the onset of polariton BEC. Polariton distributions below and just at the condensation threshold typically demonstrate higher temperature than the lattice~\cite{rodriguez2013thermalization, vakevainen2020sub, hakala2018bose, satapathy2022thermalization}
which is due to insufficient thermalization rate with respect to fast polariton decay.
Therefore, understanding polariton thermalization is pivotal for the physics of light-matter condensates in molecular systems. The actual microscopic mechanisms behind polariton thermalization remain an open question in the field. The current status is mostly relied on multimode mean-field theories, where thermalization is introduced through the Lindblad master equation or similar using effective thermalization constants~\cite{arnardottir2020multimode, zasedatelev2021single}. Thermalization rates are typically defined by comparison with experimental data as the best-fit parameters. Recently, Satapathy et al. have shed light on the nature of thermalization in organic microcavities, experimentally demonstrating the central role of the emission Stokes shift of molecules in polariton thermalization towards the BEC~\cite{satapathy2022thermalization}. However, the microscopic mechanisms of polariton thermalization in molecular systems remain largely unexplored. 

The material properties of molecular systems significantly influence thermalization behavior of polaritons~\cite{rodriguez2013thermalization, vakevainen2020sub, satapathy2022thermalization, peng2023polaritonic}. In exploring the microscopic origins of thermalization, one must consider the typical energy scale between nearest neighbor polariton states $\sim0.1~{\rm meV}$. Indeed, the finite size of a realistic system leads to the quantization of polariton states separating them in energy by $\Delta\omega_{\rm min}\sim S^{-1}$~\cite{deng2010exciton} (see Section 3 for details). This energy scale is inconsistent with photon emission from excited states ($\sim1000~{\rm meV}$) and with strong high-energy molecular vibrations ($\sim100~{\rm meV}$). Although coupling to high-energy molecular vibrations enables an efficient energy relaxation mechanism in organic polariton systems~\cite{coles2011vibrationally} driving them towards polariton BEC~\cite{zasedatelev2019room, zasedatelev2021single}, it is unlikely to be the thermalization mechanism within the lower polariton branch. However, in addition to high-energy vibrations, molecules, especially in densely packed molecular layers, such as polymer films, exhibit a wide range of low-energy vibrational modes~\cite{johnston2003low, shi2022small}. The same type of exciton-vibration coupling between electronic states and low-energy molecular vibrations can bridge nearest neighbor polariton states, matching this energy scale very efficiently. Similar mechanism of electron-acoustic phonon interaction is known to be important for polariton thermalization in crystalline semiconductor microcavities bearing Wannier-Mott type excitons~\cite{deng2010exciton,kavokin2017microcavities}. 

Direct observation of low-frequency vibrational modes below 10~meV ($\lesssim100~cm^{-1}$) in most spectroscopic experiments is challenging. Nonetheless, these modes are omnipresent and influence the spectral properties of molecules implicitly as well as the dynamics of excited states~\cite{bardeen1997temperature, karabunarliev2001franck}.
This influence is evident in phenomena such as the Stokes shift of the $0-0$ vibronic emission peak relative to absorption~\cite{karabunarliev2001franck} or temperature-dependent broadening of the emission spectrum~\cite{bardeen1997temperature}. 
Being identified as torsional and librational degrees of freedom of conjugated rings at a molecular backbone~\cite{bredas2004charge, beljonne2005excitation, tretiak2002conformational, karabunarliev2001franck} as well as longitudinal acoustic modes
of the backbone~\cite{karabunarliev2001franck, gierschner2003optical, wiesenhofer2006molecular}, low-energy vibrations constitute one of the first relaxation processes that take place within approximately 10~fs after photoexcitation to excited electronic states, known as structural (or geometric) relaxation~\cite{tretiak2002conformational, beenken2004spectroscopic, hennebicq2006chromophores}. This relaxation, akin to high-energy molecular vibrations, is driven by exciton-vibration coupling~\cite{bardeen1997temperature, karabunarliev2001franck, tretiak2002conformational}. Considering that any subsequent relaxation or deactivation of electronic excitations typically follows this geometric relaxation~\cite{hennebicq2006chromophores} and given the short lifetime of polaritons in molecular systems, we propose that low-energy vibrations are one of the primary candidates for polariton thermalization.

In this work, we developed a microscopic theory for polariton thermalization via low-energy molecular vibrations coupled to electronic degrees of freedom and derived a simple analytical expression based on the Stokes shift and linewidth of the 0-0 vibronic peak in emission spectra. We calculated the polariton thermalization rate in a practical microcavity structure and provided a recipe for cavity design and the choice of molecular system to achieve the desired thermalization rate. Last but not least, we revealed important temperature dependence that provides a quantitative understanding of the observed thermalization bottleneck effect at low temperature~\cite{plumhof2014room}.

\section{Hamiltonian of a molecular system with strong light-matter and vibronic interactions}

 \label{sec:hamiltonian}

In this Section, we develop a microscopic model to describe molecules that exhibit strong exciton-vibration (vibronic) interactions and are coupled to an optical cavity, e.g. the class of organic microcavities. We start our description with electronic and vibrational degrees of freedom of a molecular system itself, such as thin molecular films.


We consider our molecular system consisting of $N_{\rm mol}$ molecules. Each of them hosts Frenkel exciton and $N_{\rm vib}$ vibrational modes.
Strong localization of Frenkel excitons~\cite{yamamoto2003semiconductor}, enable us to define total Hamiltonian of the organic molecular film~$\hat H_{\rm mol}$ as a sum of the individual terms~$\hat H_{\rm mol}^{(m)}$
\begin{equation} \label{H_film}
\hat H_{\rm mol} = 
\sum_{m=1}^{N_{\rm mol}}
\hat H_{\rm mol}^{(m)}.
\end{equation}
The Hamiltonian of $m$th molecule is~\cite{kirton2013nonequilibrium, kirton2015thermalization, reitz2019langevin, shishkov2019enhancement}
\begin{multline} \label{H_mol}
\hat H_{\rm mol}^{(m)} = 
\hbar \omega_e^{(m)} \hat \sigma^{(m)\dag} \hat \sigma^{(m)} + 
\sum_{j=1}^{N_{\rm vib}} \hbar \omega_{{\rm v}j}^{(m)} \hat b^{(m)\dag}_{j} \hat b_{j}^{(m)} + \\ 
\sum_{j=1}^{N_{\rm vib}} \hbar \Lambda_j^{(m)} \omega_{{\rm v}j}^{(m)} \hat \sigma^{(m)\dag} \hat \sigma^{(m)} (\hat b^{(m)\dag}_j + \hat b_j^{(m)}) + \\ 
\hbar \Omega^{(m)} (\hat \sigma^{(m)} e^{i\omega_\Omega t} + \hat \sigma^{(m)\dag} e^{-i\omega_\Omega t}),
\end{multline}
where $\omega_e^{(m)}$ is the energy of the exciton of the $m$th molecule, $\omega_{{\rm v}j}^{(m)}$ is the eigenfrequency of the $j$th vibrational mode of the $m$th molecule, $\Lambda_j^{(m)}$ is the square of Huang--Rhys factor of the $j$th vibrational mode of the $m$th molecule, $\hat \sigma^{(m)}$ ($\hat \sigma^{(m)\dag}$) is the annihilation (creation) operator of the exciton, $\hat b_j^{(m)}$ ($\hat b_j^{(m)\dag}$) is the annihilation (creation) operator of the $j$th vibrational mode, $\Omega^{(m)}$ is the interaction constant between the molecule and the incident light with the frequency $\omega_\Omega$.
The parameters for each molecule is slightly different due to inherent disorder of organic systems~\cite{bassler1999site, schindler2004universal, hoffmann2010determines}. 

Given the strong exciton-vibration interaction in organic molecules that may exceed vibrational eigenfrequencies ~\cite{karabunarliev2001franck, kahle2018interpret, bassler1999site, kirton2013nonequilibrium, kirton2015thermalization}, we transition to the dressed exciton and vibrational operators.
\begin{equation} \label{sigma2S}
\hat \sigma^{(m)} = \hat S^{(m)} e^{-\sum_{j=1}^{N_{\rm vib}} \Lambda_j^{(m)} (\hat B_j^{(m)\dag} - \hat B_j^{(m)})},
\end{equation}
\begin{equation} \label{b2B}
\hat b_j^{(m)} = \hat B_j^{(m)} - \Lambda_j^{(m)} \hat S^{(m)\dag} \hat S^{(m)},
\end{equation}
where operators $\hat S^{(m)}$ and $\hat B_j^{(m)}$ are the annihilation operators of the dressed excitons and dressed molecular vibrations. 
This operators fulfill the following commutation relations $[\hat S^{(m)}, \hat S^{(m)\dag}]=[\hat \sigma^{(m)}, \hat \sigma^{(m)\dag}]$, $[\hat B_j^{(m)\dag}, \hat B_{j'}^{(m)\dag}]=[\hat b_j^{(m)}, \hat b_{j'}^{(m)\dag}]$, $[\hat S^{(m)}, \hat B_j^{(m)}]=0$ and $[\hat S^{(m)}, \hat B_j^{(m)\dag}]=0$.
The transition to the dressed operators~(\ref{sigma2S})--(\ref{b2B}) diagonalizes the part of Hamiltonian~(\ref{H_mol}) corresponding to exciton-vibration interaction and bring the Hamiltonian to the following form
\begin{multline} \label{H_mol_sB}
\hat H_{\rm mol}^{(m)} = 
\hbar \omega_0^{(m)} \hat S^{(m)\dag} \hat S^{(m)} + 
\sum_{j=1}^{N_{\rm vib}}\hbar \omega^{(m)}_{{\rm v}j} \hat B_j^{(m)\dag} \hat B_j^{(m)} 
+ 
\\ 
\hbar \Omega^{(m)} 
\left(
\hat S^{(m)} \prod_{j=1}^{N_{\rm vib}} \hat D_j^{(m)} e^{i\omega_\Omega t}
+
h.c.
\right) ,
\end{multline}
where we introduce the displacement operator 
\begin{equation} \label{displacement}
\hat D_j^{(m)} = e^{-\Lambda_j^{(m)} (\hat B_j^{(m)\dag} - \hat B_j^{(m)})},
\end{equation}
and the energy of the vibrationally dressed excitons
\begin{equation}
\omega_0^{(m)} = \omega_e^{(m)} - \sum_{j=1}^{N_{\rm vib}} \left( \Lambda_j^{(m)} \right)^2 \omega_{{\rm v}j}^{(m)}.
\end{equation}

In the next step, we incorporate the modes of the electromagnetic field inside the cavity into the system.
We also replace the energy of the dressed excitons, $\omega_0^{(m)}$, dressed vibrations, $\omega_{{\rm v}j}^{(m)}$, and their interaction constant, $\Lambda_j^{(m)}$, of each individuall molecule by the mean values, $\omega_0$, $\omega_{{\rm v}j}$, and $\Lambda_j$.

The cavity modes possess different transverse (in-plane) momenta, represented by $\hbar {\bf k}_{\parallel}$ (hereinafter referred as $\hbar{\bf k}$). 
Consequently, the full Hamiltonian of the systems now reads:
\begin{multline}\label{FullHamiltonian}
\hat H = 
\hat H_{\rm mol}
+
\sum\limits_{\bf k} 
\hbar {\omega _{{\rm cav}{\bf k}}}
\hat a_{{\rm cav}{\bf k}}^\dag {{\hat a}_{{\rm cav}{\bf k}}}
+
\\
\sum\limits_{m=1}^{N_{\rm mol}} 
\sum\limits_{{\bf k}} 
{\hbar \Omega_{\bf k}^{(m)}}
\left( 
\hat \sigma^{(m)\dag}  {{\hat a}_{{\rm{cav}}{\bf{k}}}}{e^{i{\bf k}{\bf r}^{(m)}}} 
+
h.c.
\right),
\end{multline} 
where $\hat a_{{\rm{cav}}{\bf{k}}}^\dag$ (${\hat a_{{\rm{cav}}{\bf{k}}}}$) is the creation (annihilation) operator of a photon in the cavity with the wave vector ${\bf k}$ and frequency $\omega_{{\rm{cav}}{\bf{k}}}$, which obey bosonic commutation relation ${ \left[ \hat a_{{\rm cav}{\bf k}}, \hat a_{{\rm cav}{\bf k'}}^\dag \right] = \delta_{{\bf k},{\bf k'}} }$.
Here, we assume that electric field of the $\bf{k}$th mode is distributed in a plane parallel to the mirrors according to $e^{i \bf{kr}}$~\cite{zasedatelev2019room, plumhof2014room, zasedatelev2021single, scafirimuto2021tunable, jiang2022exciton, mcghee2021polariton, mcghee2022polariton}.
Vector ${\bf r}^{(m)}$ points to the position of the $m$th  molecule. 
The single molecule light-matter interaction energy is $\hbar\Omega_{\bf k}^{(m)} =  - {{\bf E}_{\bf k}^{(m)}}{\bf d}^{(m)}$~\cite{scully1997quantum}, where ${\bf d}^{(m)}$ is the transition dipole moment of the molecule, and ${\bf E}_{\bf k}^{(m)}$ represents electric field amplitude for ``one photon'' in the cavity with the in-plane momentum $\hbar {\bf k }$ at the $m$th molecule position.


Expanding upon operators for the dressed excitons given in Eq.~(\ref{sigma2S}) and the vibrational displacement in Eq.~(\ref{displacement}) using the approximation $e^{-\Lambda_j(\hat B_j^{(m)\dag}-\hat B_j^{(m)})} \approx 1 - \Lambda_j(\hat B_j^{(m)\dag}-\hat B_j^{(m)})$ we bring the total Hamiltonian of the system, as shown in Eq.~(\ref{FullHamiltonian_dressed_approx}) to the following form 

\begin{multline}\label{FullHamiltonian_dressed_approx}
\hat H = 
\sum\limits_{\bf{k}} 
\hbar {\omega _{{\rm cav}{\bf k}}}
\hat a_{{\rm cav}{\bf k}}^\dag {{\hat a}_{{\rm cav}{\bf k}}}
+
\\
\sum_{m=1}^{N_{\rm mol}} 
\hbar \omega_0 
\hat S^{(m)\dag} \hat S^{(m)}
+
\sum_{j=1}^{N_{\rm vib}} 
\sum_{m=1}^{N_{\rm mol}} 
\hbar \omega_{{\rm v}j}\hat B_j^{(m)\dag}\hat B_j^{(m)} 
+
\\
\sum\limits_{m=1}^{N_{\rm mol}} 
\sum\limits_{{\bf k}} 
\hbar \Omega_{\bf k}^{(m)}
\left( 
\hat S^{(m)\dag}{{\hat a}_{{\rm cav}{\bf k}}}{e^{i{\bf k}{\bf r}^{(m)}}} 
+
h.c.
\right)
-
\\
\sum_{j=1}^{N_{\rm vib}}
\sum\limits_{m=1}^{N_{\rm mol}} 
\sum\limits_{{\bf k}} 
\hbar \Lambda_j \Omega _{\bf k}^{(m)}
\left( 
\hat S^{(m)\dag}
\hat B_j^{(m)} 
{{\hat a}_{{\rm cav}{\bf k}}}{e^{i{\bf k}{\bf r}^{(m)}}}
+
h.c.
\right)
+
\\
\sum_{j=1}^{N_{\rm vib}}
\sum\limits_{m=1}^{N_{\rm mol}} 
\sum\limits_{{\bf k}} 
\hbar \Lambda_j \Omega _{\bf k}^{(m)}
\left( 
\hat S^{(m)\dag}
\hat B_j^{(m)\dag}
{{\hat a}_{{\rm cav}{\bf k}}}{e^{i{\bf k}{\bf r}^{(m)}}}
+
h.c.
\right)
.
\end{multline}
We restrict ourselves to the negative exciton-photon detuning; the case of positive detuning can be considered similarly.

It is convenient to introduce the collective operators effectively describing phase coherent excitonic and vibrational states and the Rabi frequency
\begin{equation} \label{ExcitonsFourier}
 \hat c_{{\rm exc}{\bf k}} 
 = 
\frac{1 }{ \Omega_R}
\sum_{m=1}^{N_{\rm mol}}  
 \Omega_{\bf k}^{(m)}
 \hat S^{(m)} e^{i{\bf k}{\bf r}^{(m)}},
 \end{equation}
 \begin{equation} \label{VibronsFourier}
 {\hat c_{j{\bf{k}}}} = 
 \frac{1}{\sqrt{N_{\rm mol}}}
\sum_{m=1}^{N_{\rm mol}} 
 \hat B_j^{(m)}{e^{i{\bf k}{\bf r}^{(m)}}},
 \end{equation}
 \begin{equation} \label{Rabi}
\Omega_R = \sqrt{\sum_{m=1}^{N_{\rm mol}} |\Omega_{\bf k}^{(m)}|^2}.
\end{equation}
The operators~(\ref{VibronsFourier}) obey bosonic commutation relations  $\left[ \hat c_{l{\bf{k}}},\hat c_{l'{\bf{k'}}}^\dag  \right] = {\delta _{l,l'}\delta _{{\bf{k}},{\bf{k'}}}}$.
In the limit of large $N_{\rm mol}$ and small enough excitation density, the operators for excitons~(\ref{ExcitonsFourier}) exhibit bosonic properties, i.e. $\left[ \hat c_{{\rm exc}{\bf k}},\hat c_{{\rm exc}{\bf k'}}^\dag  \right] = \delta_{{\bf k},{\bf k}'}$~\cite{combescot2008microscopic}.
This approximation is consistent with the most experiments ~\cite{zasedatelev2019room, zasedatelev2021single} where the number of molecules in the illuminated area is $\sim 10^8$ and the occupation of the excitons does not surpass $\sim 0.1$.
The Hamiltonian of the system Eq.~(\ref{FullHamiltonian_dressed_approx}), in terms of the collective operators, reads:
\begin{multline}\label{FullHamiltonian_bright}
\hat H = 
\sum\limits_{\bf{k}} 
\hbar \omega_{{\rm cav}{\bf k}}
\hat a_{{\rm cav}{\bf k}}^\dag \hat a_{{\rm cav}{\bf k}}
+
\sum\limits_{\bf{k}} 
\hbar \omega_0 
\hat c_{{\rm exc}{\bf k}}^\dag \hat c_{{\rm exc}{\bf k}}
+
\\
\sum_{j=1}^{N_{\rm vib}}
\sum\limits_{\bf k} 
\hbar \omega_{{\rm v}j}\hat c_{j{\bf k}}^\dag \hat c_{j{\bf k}}
+
\sum\limits_{\bf k} 
\hbar \Omega_R
\left( 
\hat c_{{\rm exc}{\bf k}}^\dag \hat a_{{\rm cav}{\bf k}} 
+ 
h.c.
\right)
-
\\
\sum_{j=1}^{N_{\rm vib}}
\sum\limits_{{\bf k}, {\bf k}'} 
\frac{\hbar \Lambda_j \Omega_R }{ \sqrt{N_{\rm mol}}}
\left( 
\hat c_{{\rm exc}{\bf k' }}^\dag
\hat c_{j{\bf k' }-{\bf k }}
\hat a_{{\rm cav}{\bf k }}
+
h.c.
\right)
+
\\
\sum_{j=1}^{N_{\rm vib}}
\sum\limits_{{\bf k}, {\bf k}'} 
\frac{\hbar \Lambda_j \Omega_R }{ \sqrt{N_{\rm mol}}}
\left( 
\hat c_{{\rm exc}{\bf k' }}^\dag
\hat c_{j{\bf k }-{\bf k' }}^\dag
\hat a_{{\rm cav}{\bf k }}
+
h.c.
\right)
,
\end{multline} 

Considering the strong light-matter interaction with the collective vibrationally dressed excitonic states, we introduce operators for both the lower exciton-polaritons, denoted as~$\hat s_{{\rm low}{\bf k}}$, and  the upper exciton-polariton, denoted as~$\hat s_{{\rm up}{\bf k}}$, 
\begin{equation} \label{TransformationForLowerPolaritons}
\hat s_{{\rm low}{\bf k}} 
= 
\hat a_{{\rm cav}{\bf k}} \cos \varphi_{{\bf k}} 
- 
\hat c_{{\rm exc}{\bf k}} \sin \varphi_{{\bf k}}
\end{equation}
\begin{equation} \label{TransformationForUpperPolaritons}
\hat s_{{\rm up}{\bf k}} 
= 
\hat a_{{\rm cav}{\bf k}} \sin \varphi_{{\bf k}} 
+
\hat c_{{\rm exc}{\bf k}} \cos \varphi_{{\bf k}},
\end{equation}

These polariton operators describe delocalized light-matter states characterized by the in-plane momentum $\hbar \textbf{k}$. Note that, despite disorder in molecular systems, the value $\textbf{k}$ is a good quantum number, at least in strongly coupled organic microcavities. Indeed, the bottleneck effect in polariton thermalization, observed experimentally at low temperatures~\cite{plumhof2014room}, underscores the crucial role of the density of states change in the vicinity of the inflection point in the momentum space~\cite{tassone1997bottleneck}.
Angle-resolved experiments on non-ground state condensation further demonstrate the in-plane momentum conservation~\cite{baranikov2020all}. The importance of a proper quantization of states in the momentum space is also evident in recent single-photon stimulated experiments~\cite{zasedatelev2021single}, where the momentum distribution of a seed beam has to be matched to the size of the system in order to provide nonlinearity at the single photon level.

Now, we can transform our Hamiltonian in Eq.~(\ref{FullHamiltonian_bright}) into the basis of the exciton-polariton states
\begin{multline}\label{FullHamiltonian_polaritons}
\hat H = 
\sum\limits_{\bf k} 
\hbar \omega_{{\rm low}{\bf k}}
\hat s_{{\rm low}{\bf k}}^\dag \hat s_{{\rm low}{\bf k}}
+ 
\sum\limits_{\bf k} 
\hbar \omega_{{\rm up}{\bf k}}
\hat s_{{\rm up}{\bf k}}^\dag \hat s_{{\rm up}{\bf k}}
+
\\
\sum_{j=1}^{N_{\rm vib}}
\sum\limits_{\bf k} 
\hbar \omega_{{\rm v}j}
\hat c_{j{\bf k}}^\dag {\hat c}_{j{\bf k}}
-
\\
\sum_{j=1}^{N_{\rm vib}}
\sum\limits_{{\bf k}, {\bf k}'} 
\frac{ \hbar \Lambda_j \Omega_R }{ \sqrt{N_{\rm mol}} }
\left[
\left(
\cos\varphi_{\bf k'}\hat s_{{\rm up}{\bf k' }}^\dag-
\sin\varphi_{\bf k'}\hat s_{{\rm low}{\bf k' }}^\dag
\right)
\right.
\\
\hat c_{j{\bf k'}-{\bf k }}
\left(
\cos \varphi_{\bf k}
\hat s_{{\rm low}{\bf k }}
+
\sin \varphi_{\bf k'}
\hat s_{{\rm up}{\bf k }}
\right)
+
h.c.
\Big]
+
\\
\sum_{j=1}^{N_{\rm vib}}
\sum\limits_{{\bf k}, {\bf k}'} 
\frac{ \hbar \Lambda_j \Omega_R }{ \sqrt{N_{\rm mol}} }
\left[
\left(
\cos\varphi_{\bf k'}\hat s_{{\rm up}{\bf k' }}^\dag-
\sin\varphi_{\bf k'}\hat s_{{\rm low}{\bf k' }}^\dag
\right)
\right.
\\
\left.
\hat c_{j{\bf k}-{\bf k' }}^\dag
\left(
\cos \varphi_{\bf k}
\hat s_{{\rm low}{\bf k }}
+
\sin \varphi_{\bf k'}
\hat s_{{\rm up}{\bf k }}
\right)
+
h.c.
\right]
,
\end{multline}
where we omit the non-resonant terms.
We define the polariton eigenfrequencies as follows:
\begin{equation} \label{FrequenciesOfLowerPolaritons}
\omega _{{\rm low}{\bf k}} 
= 
\frac{\omega_0 + \omega_{{\rm cav}{\bf k}} }{ 2}
-
\sqrt{
\frac{
\left( 
\omega_0 - \omega _{{\rm cav}{\bf k}} 
\right)^2 
}{
 4
}
+
\Omega_R^2
},
\end{equation}
\begin{equation} \label{FrequenciesOfUpperPolaritons}
\omega _{{\rm up}{\bf k}} 
= 
\frac{\omega_0 + \omega_{{\rm cav}{\bf k}} }{ 2}
+
\sqrt{
\frac{
\left( 
\omega_0 - \omega _{{\rm cav}{\bf k}} 
\right)^2 
}{
 4
}
+
\Omega_R^2
}.
\end{equation}
The ratio between the excitonic (material) and photonic parts of polariton states are defined by the Hopfield coefficients $\sin\varphi_{\bf k}$ and $\cos\varphi_{\bf k}$~\cite{kavokin2017microcavities}, where $\varphi_{\bf k}$ is
\begin{equation} \label{angle}
\varphi_{\bf k} = 
\frac{1 }{ 2}
{\rm arctg}
\left(       
\frac{2 \Omega_R}{\omega_0 - \omega_{{\rm cav}{\bf k}}}
 \right).
\end{equation}

We note that the lower polaritons inherit from cavity photons the quadratic dispersion in the vicinity of the ground state, $\omega_{{\rm low}{\bf k}}\approx\omega_{{\rm low}{\bf 0}}+\alpha_{\rm pol}{\bf k}^2$, where 
\begin{equation} \label{dispersion}
\alpha_{\rm pol}
\approx
\frac{\alpha_{\rm cav}}{2}
\left[
1
+
\frac
{
\omega_0 - \omega_{{\rm cav}{\bf k}={\bf 0}}
}
{
\sqrt{(\omega_0 - \omega_{{\rm cav}{\bf 0}})^2 + 4\Omega_R^2}
}
\right]
.
\end{equation}


As evident from the Hamiltonian in its final form given by Eq.~(\ref{FullHamiltonian_polaritons}), polaritons are the hybrid light-matter states that include matter components featuring both electronic and vibrational degrees of freedom. 
In the next section, we show that the nonlinear interplay between them gives rise to polariton thermalization through low-energy molecular vibrations.

\section{Thermalization rate of polaritons}

Due to the complexity of the vibrational landscape with its various degrees of freedom we treat low-frequency vibrations through introducing a reservoir and exclude them using Born--Markov approximation. To justify this assumption, we compare the degrees of freedom of the system with those of the reservoir. The Born–Markov approximation is applicable for the reservoir having significantly more degrees of freedom than the system itself~\cite{ferreira2021collapse, tatarskiui1987methodological}.
This scenario is precisely what occurs in our system.
Indeed, the number of states for lower polaritons roughly matches the number of states in the corresponding cavity. For instance, a cavity with its fundamental mode characterized by a wavelength $\lambda$ has a number of states $N_{\rm states} \approx \pi k_{\rm max}^2S/(2 \pi)^2$, where $S$ is the area of interest~\cite{deng2010exciton}.
In room temperature BEC experiments~\cite{zasedatelev2021single, zasedatelev2019room, plumhof2014room} 
these parameters are typically the following: $\lambda \approx 500~{\rm nm}$ and $S \approx 500~{\rm \mu m}^2$, which result in $N_{\rm states} \approx 10^4$ number of states.
Furthermore, the corresponding number of states within the energy range $k_BT$ at the ground polariton state is about $10^3$.
The number of molecules / molecular chains is good estimate for the total number of molecular vibrations, which is around $10^{8}$ within the region of interest.
In addition, we would like to highlight some empirical evidences in support to the Born--Markov approximation. For example, the thermal distribution of polaritons above the BEC threshold~\cite{plumhof2014room} 
with the temperature close to the environment is a signature of at least partial thermal equilibrium. This observation aligns with the expectations of the Born-Markov approximation.

The Hamiltonian~(\ref{FullHamiltonian_polaritons}) allows us to derive the contribution of the molecular vibrations to the thermalization rate of polaritons. 
We separate all the vibrational modes of the molecules into two groups: high-frequency vibrations and low-frequency vibrations.
We order all the vibrational modes such that $\omega_{{\rm v}n}<\omega_{{\rm v}n+1}$ and denote $M$ as the number of the vibrational mode for which $\omega_{{\rm v}M} < \Gamma$ and $\omega_{{\rm v}M+1} > \Gamma$, where $\Gamma$ is the standard deviation of dressed excitons transition frequencies.
All the vibrational modes with the natural frequency below and equal $\omega_{{\rm v}M}$ we call low-frequency vibrations, the rest vibrational modes we call high-frequency vibrations.
We take into account the low-frequency vibrations effectively, as a reservoir, while we consider high-frequency vibrations explicitly.
Thus, from Hamiltonian~(\ref{FullHamiltonian_polaritons}), we obtain the thermalization rate between two arbitrary lower polariton states with the wave vectors $\bf k$ and ${\bf k}'$ ($\omega_{{\rm low}{\bf k}} > \omega_{{\rm low}{\bf k}'}$)
\begin{multline} \label{therm rate 1}
\gamma_{\rm therm}^{{\bf k}'{\bf k}}
=
2\pi
\sin^2(\varphi_{{\bf k'}} - \varphi_{{\bf k}})
\\
\frac{
\Omega_R^2
}{
N_{\rm mol}
}
\Lambda^2(\Delta\omega_{{\bf kk}'})\nu(\Delta\omega_{{\bf kk}'})
\left(
1 + n_{\rm v}(\Delta\omega_{{\bf kk}'})
\right)
,
\end{multline}
\begin{multline} \label{therm rate 2}
\gamma_{\rm therm}^{{\bf k}{\bf k}'}
=
2\pi
\sin^2(\varphi_{{\bf k'}} - \varphi_{{\bf k}})
\\
\frac{
\Omega_R^2
}{
N_{\rm mol}
}
\Lambda^2(\Delta\omega_{{\bf kk}'})\nu(\Delta\omega_{{\bf kk}'})
n_{\rm v}(\Delta\omega_{{\bf kk}'})
,
\end{multline}
where we use the continuous limit of the distribution of the low-frequency vibrations. $\nu(\omega)$ is the density of the states of the molecular vibrations at frequency $\omega$, $\Lambda(\omega)$ is the square of Huang--Rhys factor at the frequency $\omega$,
\begin{equation} \label{n_v}
n_{\rm v}(\Delta \omega)
=
\frac{1}{e^{\hbar\Delta\omega/k_BT}-1}
,
\end{equation}
\begin{equation} \label{frequency_difference}
\Delta\omega_{{\bf kk}'} 
= 
|\omega_{{\rm low}{\bf k}} - \omega_{{\rm low}{\bf k}'}|
.
\end{equation}
The frequency difference $\Delta\omega_{{\bf kk}'}$ is limited from below due to the finite size of the system.
In general, the finite size of the system leads to the discrete spectrum~\cite{lang1973laser}.
For example, in 2D organic microcavities, the minimal value of $\Delta\omega_{{\bf kk}'}$ is reversely proportional to the area of the system~\cite{deng2010exciton}, $S$, and equal
\begin{equation} \label{omega_min}
\Delta\omega_{\rm min} = \frac{4\pi\alpha_{\rm pol}}{S}
\end{equation}
as long as we can approximate the real dispersion of the lower polaritons with the quadratic one~\cite{shishkov2022analytical}.
We note, that this procedure is independent of the correlation length of the polaritons.
Indeed, the correlation radius is determined by the density matrix of the polaritons~\cite{shishkov2022analytical, pitaevskii2016bose}.
Moreover, even for a very large system, the correlation length can be small compared to the size of the system below the condensation threshold and infinite above the condensation~\cite{shishkov2022analytical, mouchliadis2008first, pitaevskii2016bose}.

The thermalization rates (\ref{therm rate 1})--(\ref{therm rate 2}) obey the Kubo--Martin--Schwinger relation~\cite{kubo1957statistical}
\begin{equation}
\gamma_{\rm therm}^{{\bf k}'{\bf k}} 
= 
\gamma_{\rm therm}^{{\bf k}{\bf k}'} 
\exp 
\left( 
\frac{ \hbar\omega_{{\rm low}{\bf k}} - \hbar\omega_{{\rm low}{\bf k}'} }{ k_BT } 
\right)
.
\end{equation}
The corresponding Lindblad operator~\cite{breuer2002theory, zasedatelev2021single} for the density matrix of polaritons $\hat\rho$
\begin{multline} \label{Lindblad for polaritons}
L(\hat \rho) =
\sum_{{\bf k}_1, {\bf k}_2} 
\frac{\gamma_{\rm therm}^{{\bf k}_1{\bf k}_2}}{2} 
\left(
2\hat s_{{\rm low}{\bf k }_2} \hat s_{{\rm low}{\bf k }_1}^\dag 
\hat \rho
\hat s_{{\rm low}{\bf k }_1} \hat s_{{\rm low}{\bf k }_2}^\dag
-
\right.
\\
\hat s_{{\rm low}{\bf k }_1} \hat s_{{\rm low}{\bf k }_2}^\dag
\hat s_{{\rm low}{\bf k }_2} \hat s_{{\rm low}{\bf k }_1}^\dag 
\hat \rho
-
\\
\left.
\hat \rho
\hat s_{{\rm low}{\bf k }_1} \hat s_{{\rm low}{\bf k }_2}^\dag
\hat s_{{\rm low}{\bf k }_2} \hat s_{{\rm low}{\bf k }_1}^\dag 
\right)
\end{multline}
describes the energy flow from the lower polaritons having wave vector~${\bf k}_2$ towards ones with wave vector~${\bf k}_1$.

Equations (21) and (22) show that the thermalization rates, denoted as $\gamma_{\text{therm}} \sim \Lambda^{2} = S_{H}$, are directly proportional to the Huang-Rhys parameter ($S_{H}$). This parameter determines the strength of the exciton-vibration interaction and establish connection to spectroscopic properties of bare molecules. Interestingly, a Hamiltonian of the same type is also used to describe Raman scattering~\cite{shishkov2019enhancement, shishkov2021enhancement},
as well as the transition process from bright excitons to lower polaritons, which occurs through high-frequency vibrations~\cite{zasedatelev2021single,shishkov2023mapping}. Indeed, the cross-section for Raman scattering is proportional to $S_{H}$. Therefore, polariton thermalization via low-frequency vibrations predominantly involves Raman-allowed transitions. Whether these vibrations are also IR-allowed is not particularly important for this mechanism.

Given the ultrafast timescale of geometric relaxation in the electronically excited states of highly conjugated molecular systems~\cite{tretiak2002conformational, beenken2004spectroscopic, hennebicq2006chromophores}, the low-frequency vibrations emerge as the primary candidates for polariton thermalization in organic cavities. 
This thermalization process conserves the total number of polaritons, but does not preserve the total energy~\cite{shishkov2022exact, shishkov2022analytical}.
As a polariton transitions to a state with lower in-plane momentum, it loses some energy. This released energy is resonantly absorbed by molecules through the excitation of low-frequency molecular vibrations exhibiting the highest exciton-vibration coupling, which, in turn, dictates the net polariton thermalization rate.

The thermalization rates~(\ref{therm rate 1}) and (\ref{therm rate 2}) depend on the properties of the molecular vibrations with particular frequency, namely $\Lambda^2(\omega)$ and $\nu(\omega)$, the angles $\varphi_{\bf k}$ and $\varphi_{\bf k'}$ (see Eq.~(19)), and the frequency difference between them $\Delta\omega_{{\bf kk}'} = |\omega_{{\rm low}{\bf k}} - \omega_{{\rm low}{\bf k}'}|$.
To estimate the thermalization rates~(\ref{therm rate 1}) and (\ref{therm rate 2}) we should consider the properties of both molecular vibrations and polaritons.
For polariton states with wave vectors $\bf k$ and $\bf k'$ for which $\Delta\omega_{{\bf kk}'},\Delta\omega_{{\bf k0}}  \ll |\omega_{\rm exc}-\omega_{{\rm low}{\bf 0}}|$, we can expand $\sin^2(\varphi_{{\bf k'}} - \varphi_{{\bf k}})$ to the following approximate equation 
\begin{equation}
\sin^2(\varphi_{{\bf k'}} - \varphi_{{\bf k}}) 
\approx
\frac
{
\alpha_{\rm cav}^2\Omega_R^2 (\Delta \omega_{\bf k' k})^2
}{
\alpha_{\rm pol}^2((\omega_{\rm exc}-\omega_{\rm cav})^2+4 \Omega_R^2)^2
}
\end{equation}
Although local properties of low-frequency molecular vibrations are hard to access experimentally~\cite{baderschneider2016influence, stampfl1995photoluminescence},
we can reliably estimate $\omega\Lambda^2(\omega)\nu(\omega)$ and $\omega^2\Lambda^2(\omega)\nu(\omega)$ in the low-frequency range, by their average values $A_1/\omega_M$ and $A_2/\omega_M$, respectively, where
\begin{equation} \label{A1}
A_1 = \int_0^{\omega_{M}} \omega \Lambda^2(\omega) \nu(\omega) d\omega,
\end{equation}
\begin{equation} \label{A2}
A_2 = \int_0^{\omega_{M}} \omega^2 \Lambda^2(\omega) \nu(\omega) d\omega.
\end{equation}

It is very useful to consider Eq.~(\ref{therm rate 1})--(\ref{therm rate 2}) in high and low temperature limits separately. Inspired by the recent experimental progress in room temperature BEC and related phenomena we focus on high temperature limit as the most practical.

In the high temperature limit $k_BT\gg \hbar\omega_M$ we have $\hbar \Delta\omega_{{\bf kk}'} \ll k_BT$ and therefore can approximate $n_{\rm v}(\Delta\omega_{{\bf kk}'}) \approx k_BT/(\hbar \Delta\omega_{{\bf kk}'})$ obtaining the following thermalization rates:
\begin{multline} \label{second estimation}
\gamma_{\rm therm}^{{\bf k}{\bf k'}}
\approx
\frac
{
2\pi
\alpha_{\rm cav}^2\Omega_R^4
}{
\alpha_{\rm pol}^2((\omega_{\rm exc}-\omega_{\rm cav})^2+4 \Omega_R^2)^2
}
\\
\frac{
1
}{
N_{\rm mol}
}
\Delta\omega_{{\bf kk'}}
\Lambda^2(\Delta\omega_{{\bf kk'}})\nu(\Delta\omega_{{\bf kk'}})
\frac{k_B T}{\hbar}
,
\end{multline}
for any relation between $\omega_{{\rm low}{\bf k}}$ and $\omega_{{\rm low}{\bf k}'}$.

Then we estimate the average thermalization rate over frequencies $\Delta\omega_{{\bf kk'}}$ from $0$ to $\omega_M$ using Eq.~(\ref{A1})
\begin{multline} \label{thermalization_rate_average}
\frac{1}{\omega_M}
\int_0^{\omega_M}
\gamma_{\rm therm}^{{\bf k}{\bf k'}}
d\Delta\omega_{{\bf kk'}}
\approx
\\
\frac
{
2\pi
\alpha_{\rm cav}^2\Omega_R^4
}{
\alpha_{\rm pol}^2((\omega_{\rm exc}-\omega_{\rm cav})^2+4 \Omega_R^2)^2
}
\frac{
A_1
}{
N_{\rm mol}
}
\frac{k_B T}{\hbar \omega_M}.
\end{multline}

For 2D organic microcavities, the minimal value of $\Delta\omega_{{\bf kk}'}$ is reversely proportional to the area of the system, $S$, and equal $\Delta\omega_{\rm min} = 4\pi\alpha_{\rm pol}/S$.
The major contribution to the average thermalization rate $\gamma_{\rm therm}^{{\bf k}{\bf k'}}$ and $\gamma_{\rm therm}^{{\bf k'}{\bf k}}$ according to Eq.~(\ref{thermalization_rate_average}) corresponds to the transitions between the nearest polariton states.
\begin{multline}
\left(
\gamma_{\rm therm}^{{\bf k}{\bf k'}}
\right)_{k_BT\gg \hbar\omega_M}^{\rm est}
=
\\
\frac{\omega_M}{\Delta\omega_{\rm min}}
\left[
\frac{1}{\omega_M}
\int_0^{\omega_M}
\gamma_{\rm therm}^{{\bf k}{\bf k'}}
d\Delta\omega_{{\bf kk'}}
\right]
\end{multline}
Thus, we obtain thermalization rate at the high temperature limit
\begin{multline} \label{second estimation}
\left(
\gamma_{\rm therm}^{{\bf k}{\bf k'}}
\right)_{k_BT\gg \hbar\omega_M}^{\rm est}
=
\\
\frac
{
\alpha_{\rm cav}^2\Omega_R^4
}{
\alpha_{\rm pol}^2((\omega_{\rm exc}-\omega_{\rm cav})^2+4 \Omega_R^2)^2
}
\frac{
A_1
}{
N_{\rm mol}
}
\frac{S k_B T}{\hbar \alpha_{\rm pol}}
.
\end{multline}

In the low-temperature limit, where $k_BT \lesssim \hbar\omega_M$, we cannot neglect unity in the factor $(1 + n_{\rm v}(\Delta\omega_{{\bf kk}'}))$. Therefore, we must proceed with Eq.~(\ref{therm rate 1})--(\ref{therm rate 2}) in its complete form. Here we use average $A_2/\omega_M$ to estimate $\omega^2 \Lambda^2(\omega) \nu(\omega)$ and obtain the rates at the low temperature limit.

\begin{multline} \label{therm k'k}
\left(
\gamma_{\rm therm}^{{\bf k}'{\bf k}}
\right)_{k_BT\lesssim \hbar\omega_M}^{\rm est}
=
\frac
{
2\pi\alpha_{\rm cav}^2\Omega_R^4
}{
\alpha_{\rm pol}^2((\omega_{\rm exc}-\omega_{\rm cav})^2+4 \Omega_R^2)^2
}
\\
\frac{
A_2
}{
N_{\rm mol}
}
\frac{1 }{\omega_M}
\left(
1 + n_{\rm v}(\Delta\omega_{{\bf kk}'})
\right)
,
\end{multline}
\begin{multline} \label{therm kk'}
\left(
\gamma_{\rm therm}^{{\bf k}{\bf k}'}
\right)_{k_BT\lesssim \hbar\omega_M}^{\rm est}
=
\frac
{
2\pi\alpha_{\rm cav}^2\Omega_R^4
}{
\alpha_{\rm pol}^2((\omega_{\rm exc}-\omega_{\rm cav})^2+4 \Omega_R^2)^2
}
\\
\frac{
A_2 
}{
N_{\rm mol}
}
\frac{1}{\omega_M}
n_{\rm v}(\Delta\omega_{{\bf kk}'})
,
\end{multline}
where we set $\omega_{{\rm low}{\bf k}} > \omega_{{\rm low}{\bf k}'}$.

In the next Section, we show how the features of the emission and absorption spectra of an organic film is linked to the integral properties of low-frequency molecular vibrations~(\ref{A1})~and~(\ref{A2}).

\section{The role of low-energy vibrations in absorption and emission spectra} 
Determining the individual parameters for each of the low-frequency vibrational modes in densely-packed molecular systems even without optical cavities stands as an extremely challenging experimental problem. Luckily, ensemble averaged properties of vibrational modes and their coupling to electronic states can be obtained from emission and absorption spectra, for example by analysis of the widths and relative spectral positions of the maxima as function of temperature. In this section we provide theoretical analysis for the spectral properties of bare molecular systems (such as thin organic films) without a cavity and establish connection to $A_1$ and $A_2$ parameters that define the polariton thermalization rate.

In general, the theoretical calculation of the emission and absorption spectra of the molecular system requires the knowledge of Hermitian evolution of the organic film and the relaxation processes. 
The Hermitian evolution is given by the Hamiltonian~(\ref{H_film}), while the relaxation processes can be described within Lindblad approach by introducing the density matrix $\hat \rho$ of excitons and vibrations hosted by the molecules in the organic film.
The non-equilibrium dynamics of $\hat \rho$ is governed by the Lindblad master equation~\cite{scully1997quantum, carmichael2009open}
\begin{equation} \label{MasterEquation}
{\frac{ d\hat \rho}{dt}} 
= 
{\frac{ i}{\hbar}}{\left[ {\hat \rho}, {\hat H}_{\rm mol} \right]}
+
\sum_n L_{\hat A_n}(\hat \rho),
\end{equation}
where $L_{\hat A_n}(\hat \rho)$ is Lindblad operator, $\hat A_n$ is the relaxation operator 
\begin{equation}
L_{\hat A}(\hat \rho) 
=
\hat A \hat \rho \hat A^\dag 
- 
\frac{1 }{ 2} \hat \rho \hat A^\dag \hat A
- 
\frac{1 }{ 2} \hat A^\dag \hat A \hat \rho 
\end{equation}
We consider energy dissipation of both: electronic and vibrational subsystems of molecules. The incoherent pumping of dressed excitonic states is also introduced via Lindblad operator. Therefore, all relaxation processes and the incoherent pumping correspond to the following operators $\sqrt{\gamma_{\rm diss}^{(m)}} \hat S^{(m)}$, $\sqrt{\gamma_{\rm pump}^{(m)}} \hat S^{(m)\dag}$,  $\sqrt{2(1+ n_{{\rm v}j}^{(m)}) \gamma_{{\rm v}j}^{(m)}} \hat B_j^{(m)}$ and $\sqrt{2n_{{\rm v}j}^{(m)} \gamma_{{\rm v}j}^{(m)}} \hat B_j^{(m)\dag}$ respectively, where
\begin{equation} \label{n_vjm}
n_{{\rm v}j}^{(m)}
=
\frac{1}{e^{\hbar\omega_{{\rm v}j}^{(m)}/k_BT}-1}
.
\end{equation}
Note the dynamics of molecular vibrations require both relaxation operators $\sqrt{2(1+ n_{{\rm v}j}^{(m)}) \gamma_{{\rm v}j}^{(m)}} \hat B_j^{(m)}$ and $\sqrt{2n_{{\rm v}j}^{(m)} \gamma_{{\rm v}j}^{(m)}} \hat B_j^{(m)\dag}$ as the thermal energy $k_BT$ can exceed low-energy molecular vibrations $\hbar \omega_{{\rm v}j}^{(m)}$. 



Considering that the relative differences in the parameters, stemming from disordered nature of molecular systems, are all of comparable magnitude, the most significant absolute difference lies in $\omega_0^{(m)}$. 
Therefore, we can state that the molecules within the system are distinguishable solely by the energy of the dressed excitons. 
Given this, we assume that the dephasing of excitons in the system originates primarily from two factors: (i) inhomogeneous broadening due to the difference in the energies of the dressed excitons, and (ii) the impact of low-energy vibrations on the dynamics of the excitons. 
These factors are explicitly considered in Hamiltonian~(\ref{H_film}). And thus we do not need to account for them separately in the Lindblad operators given by Eq.~(\ref{MasterEquation}).

In light of the abovementioned, we will disregard the variations in all other molecular parameters, therefore, the index $(m)$ is omitted hereinafter as identified in Eq.~(\ref{n_vjm}).

\subsection{Emission and absorption spectra} \label{subsec: em and abs spectra}
The stationary emission spectrum of the molecular system itself can be calculated from two-time correlator $\left\langle \hat\sigma^\dag(t+\tau)\hat\sigma(t) \right\rangle$~\cite{carmichael2009open}.
The presence of the weak incoherent pumping implies the assumption: $\gamma_{\rm pump} \neq 0$, $\gamma_{\rm pump} \ll \gamma_{\rm diss}$, $\Omega = 0$ in~Eq.~(\ref{MasterEquation}), which corresponds the low probability of any of the molecules to be in the excited state~\cite{carmichael2009open}.

The absorption spectrum is determined by the imaginary part of the permittivity that characterizes linear response on the weak external field~\cite{loudon2000quantum}.
Therefore, absorption spectrum is proportional to $-\langle \hat \sigma(t)e^{i\omega_\Omega t} \rangle/\Omega$ at the condition $\gamma_{\rm pump}=0$, $\Omega \neq 0$.

Given the typical molecular density of organic thin films utilized in polariton research is $10^{20}-10^{21}~{\rm cm}^{-3}$ we deal with the continuous distribution of exciton energies.
Moreover, taking into account the disorder mentioned above it is reasonable to assume normally distributed energies with the standard deviation $\Gamma$.
We also assume that $\Gamma \gg \gamma_{\rm diss}/2 + \gamma_{{\rm v}\Sigma}$.
Thus, the inhomogeneous broadening leads to the Gaussian lineshape~\cite{knapp1984lineshapes, kador1991stochastic, spano2009determining, guha2003temperature, borrelli2014theoretical, ostroverkhova2016organic} of the spectral peaks.

Using the master equation~(\ref{MasterEquation}) we find emission and absorption spectra of the bare molecular system without a cavity (see Supplementary Information)

\begin{multline} \label{I_em_1}
I_{\rm em}(\omega) 
= 
\sum_{k'_1,...,k'_{N_{\rm vib}}} 
z_{k'_1}(n_{{\rm v}1}\Lambda_1^2)
...
z_{k'_{N_{\rm vib}}}(n_{{\rm v}N_{\rm vib}}\Lambda_{N_{\rm vib}}^2)
\\
\sum_{k_1,...,k_{N_{\rm vib}}}  
z_{k_1}((1+n_{{\rm v}1})\Lambda_1^2)...z_{k_M}((1+n_{{\rm v}N_{\rm vib}})\Lambda_{N_{\rm vib}}^2)
\\
\frac{
1
}{
\sqrt{2\pi \Gamma^2}
}
e^{ - (\omega - \omega_0 + \sum_{j=1}^{N_{\rm vib}}\omega_{{\rm v}j}k_j)^2 / 2\Gamma^2 }
,
\end{multline}
\begin{multline}  \label{I_abs_1}
I_{\rm abs}(\omega) 
= 
\sum_{k'_1,...,k'_{N_{\rm vib}}} 
z_{k'_1}(n_{{\rm v}1}\Lambda_1^2)
...
z_{k'_{N_{\rm vib}}}(n_{{\rm v}N_{\rm vib}}\Lambda_{N_{\rm vib}}^2)
\\
\sum_{k_1,...,k_{N_{\rm vib}}}  
z_{k_1}((1+n_{{\rm v}1})\Lambda_1^2)...z_{k_M}((1+n_{{\rm v}N_{\rm vib}})\Lambda_{N_{\rm vib}}^2)
\\
\frac{
1
}{
\sqrt{2\pi \Gamma^2}
}
e^{ - (\omega - \omega_0 - \sum_{j=1}^{N_{\rm vib}}\omega_{{\rm v}j}k_j)^2 / 2\Gamma^2 }
,
\end{multline}
where the sum for $k_1$, $k_2$, ..., $k_M$ runs over all non-negative integers and 
\begin{equation}
z_n(x) = \frac{x^n e^{-x} }{ n!},
\end{equation}

The expression for the absorption spectrum (Eq.~(\ref{I_abs_1})) is distinguished from that of the luminescence spectrum (Eq.~(\ref{I_em_1})) by the substitution of the sign preceding $\sum_{j=1}^{N_{\rm vib}}\omega_{{\rm v}j}k_j$. Consequently, the non-linear interaction between electronic states and low-frequency vibrations of molecular nuclei is a contributing factor to the Stokes shift \cite{karabunarliev2001franck}.

\begin{figure}
\includegraphics[width=1\linewidth]{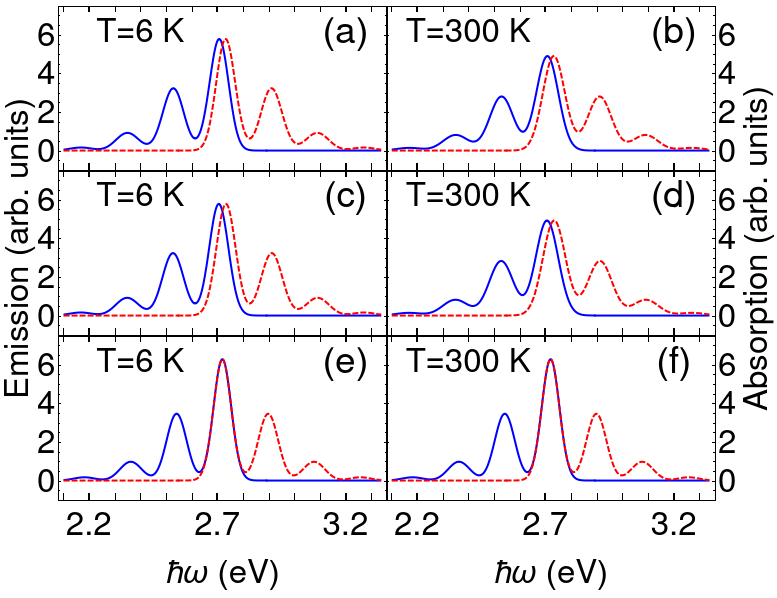}
\caption{
The emission (blue solid line) and absorption (red dashed line) spectra of a bare MeLPPP layer at different temperatures (a), (c), (e) $T=6~{\rm K}$; (b), (d), (f) $T=300~{\rm K}$.
The spectra in (a) and (b) explicitly take into account all the vibrational modes according to exact expressions Eq.~(\ref{I_em_1}) and (\ref{I_abs_1}), the spectra (c) and (d) take into account low-frequency vibrational modes effectively according to Eq.~(\ref{I em renormalized})~and~(\ref{I abs renormalized}), the spectra (e) and (f) utilize the exact model while neglecting low-frequency modes.
} \label{fig: spectra}
\end{figure}


We can effectively take into account low-frequency vibrational modes and obtain from Eq.~(\ref{I_em_1}) and (\ref{I_abs_1}) the reduced expression for emission and absorption spectra (see Supplementary Information)
\begin{multline} \label{I em renormalized}
I_{\rm em}(\omega) 
= 
\sum_{k_{M+1},...k_{N_{\rm vib}}} z_{k_{M+1}}(\Lambda_{M+1}^2) ... z_{k_{N_{\rm vib}}}(\Lambda_{N_{\rm vib}}^2)
\\
\frac{
1
}{
\sqrt{2\pi \Gamma_{\rm em}^2}
}
e^{ - (\omega - \omega_{\rm em} + \sum_{j={M+1}}^{N_{\rm vib}}\omega_{{\rm v}j}k_j)^2 / 2\Gamma_{\rm em}^2 }
,
\end{multline}
\begin{multline}  \label{I abs renormalized}
I_{\rm abs}(\omega) 
= 
\sum_{k_{M+1},...k_{N_{\rm vib}}} z_{k_{M+1}}(\Lambda_{M+1}^2) ... z_{k_{N_{\rm vib}}}(\Lambda_{N_{\rm vib}}^2)
\\
\frac{
1
}{
\sqrt{2\pi \Gamma_{\rm abs}^2}
}
e^{ - (\omega - \omega_{\rm abs} - \sum_{j={M+1}}^{N_{\rm vib}}\omega_{{\rm v}j}k_j)^2 / 2\Gamma_{\rm abs}^2 }
,
\end{multline}
where we assume $n_{{\rm v}j} \approx 0$ for high-frequency vibrations and we introduce
\begin{equation} \label{omega em renormalized} 
\omega_{\rm em} = \omega_0 - \sum_{j=1}^{M}\Lambda_j^2\omega_{{\rm v}j},
\;\;\;
\omega_{\rm abs} = 2\omega_0 - \omega_{\rm em}, 
\end{equation}
\begin{equation} \label{gamma em renormalized} 
\Gamma_{\rm em}^2 = \Gamma^2 + \sum_{j=1}^{M}\Lambda_j^2\omega_{{\rm v}j}^2 \left(1 + 2 n_{{\rm v}j} \right),
\;\;\;
\Gamma_{\rm abs} = \Gamma_{\rm em}.
\end{equation}
In the limit of the continuous distribution of the low-frequency vibrations Eq.~(\ref{omega em renormalized}) and~(\ref{gamma em renormalized}) become 
\begin{equation} \label{omega em renormalized continuous}
\omega_{\rm em} = \omega_0 - \int_0^{\omega_{M}} \omega \Lambda^2(\omega) \nu(\omega) d\omega
\end{equation}
\begin{multline} \label{gamma em renormalized continuous}
\Gamma_{\rm em}^2 
= 
\Gamma^2 
+
\int_0^{\omega_{M}} \omega^2 \Lambda^2(\omega) \left( 1 + 2n_{\rm v}(\omega) \right) \nu(\omega) d\omega
\end{multline}
here we make use of the same notations as in Eq.~(\ref{therm rate 1})--Eq.~(\ref{therm rate 2}), in particular, $n_{\rm v}(\omega)$ is determined by Eq.~(\ref{n_v}).

\begin{figure}
\includegraphics[width=1\linewidth]{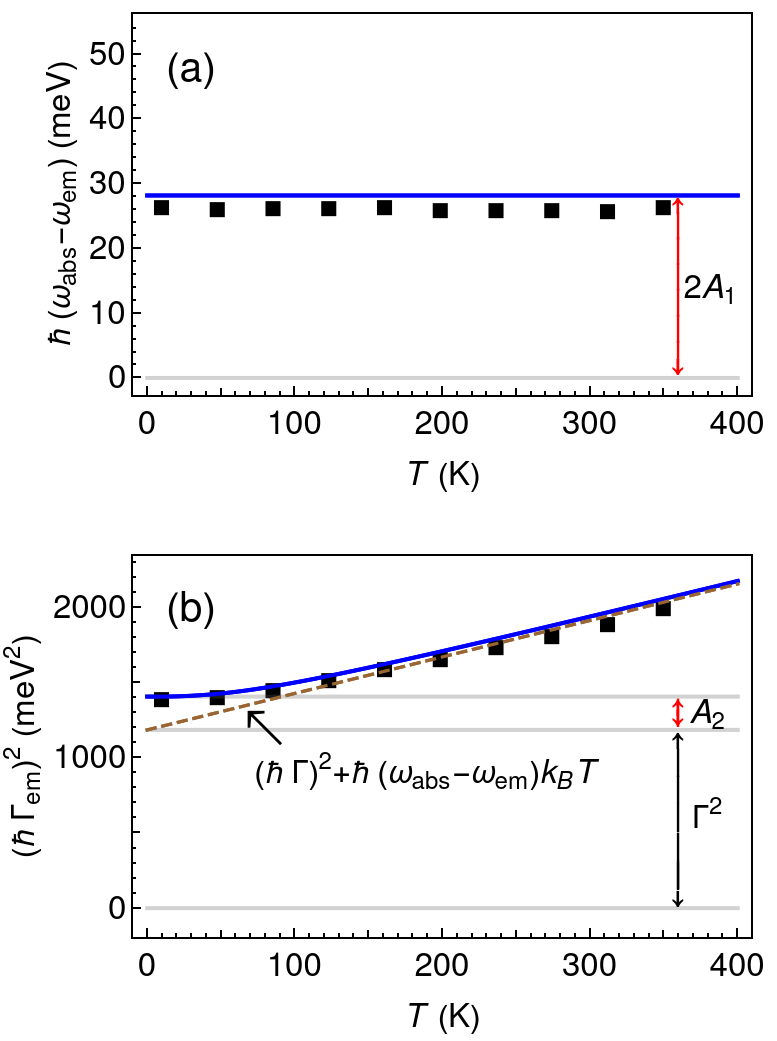}
\caption{
Temperature dependence of the Stokes shift - (a) and linewidth - (b) of $0-0$ emission peak. The dashed line shows the asymptotic behaviour of the linewidth at high temperatures.
Red double arrows mark the net values of the low-frequency vibration, namely $A_1$ and $A_2$ parameters.
} \label{fig: illustration}
\end{figure}

\subsection{Stokes shift}

From Eq.~(\ref{I em renormalized})--(\ref{gamma em renormalized}), it is evident that, at a given temperature, the line shape of the emission spectrum does not enable us to differentiate between inhomogeneous broadening and the effects of low-frequency vibrations. However, a combined analysis of both the emission and absorption spectra allows for this distinction. Unlike low-frequency vibrations, inhomogeneous broadening does not cause the Stokes shift between 0-0 vibronic peaks in the absorption $\omega_{\rm abs}$ and emission $\omega_{\rm em}$. Therefore, the Stokes shift, $\omega_{\rm abs} - \omega_{\rm em}$, provides insights into the low-frequency vibrational dynamics.

Indeed, from Eq.~(\ref{omega em renormalized})~and~(\ref{omega em renormalized continuous}) one can obtain the Stokes shift:
\begin{equation} \label{integral 1}
\omega_{\rm abs}-\omega_{\rm em} = 2\int_0^{\omega_{M}} \omega \Lambda^2(\omega) \nu(\omega) d\omega,
\end{equation}

The Stokes shift is nearly independent of temperature. Therefore, analysis of the emission and absorption spectra of the molecular system itself (without a cavity) enables us to infer the net properties of low-frequency vibrations -- essential for the polariton thermalization rate as delineated in Eq.~(\ref{A1}).

\subsection{Temperature-dependent broadening}
The second integral property of the low-frequency vibrations can be determined from the temperature dependence of the linewidth of the $0-0$ vibronic peak in the emission spectrum. At high temperatures, $k_BT \gtrsim \hbar \omega_M$, expression for the linewidth Eq.~(\ref{gamma em renormalized continuous}) takes the following form:
\begin{equation} \label{Gamma high temperature}
\Gamma_{\rm em}^2
=
\Gamma^2
+
\frac{k_BT }{ \hbar} 
(\omega_{\rm abs}-\omega_{\rm em}).
\end{equation}
The obtained equation above not only outlines the temperature dependence of the linewidth for the 0-0 vibronic peak with respect to the Stokes shift but also emphasizes the self-consistency
of the theory presented. 
Furthermore, Eq.~(\ref{Gamma high temperature})  provides the means to calculate the degree of inhomogeneous broadening denoted by $\Gamma^2$.


At the low-temperature limit, the linewidth of the $0-0$ emission peak~(\ref{omega em renormalized continuous}) approaches a constant value.
\begin{equation} \label{Gamma low temperature}
\Gamma_{\rm em}^2
=
\Gamma^2
+
\int_0^{\omega_{M}} \omega^2 \Lambda^2(\omega) \nu(\omega) d\omega
\end{equation}
Given the value of $\Gamma^2$ is extracted from the analysis at high temperatures limit discussed above, using Eq.~(\ref{Gamma low temperature}) we can now obtain the second net property of low-frequency vibrations~(\ref{A2}) that define polariton thermalization rate.

\section{Calculation of the thermalization rate in an organic microcavity} \label{sec: example}
Here we focused on organic Fabry--Pérot microcavities as the prototype system for these calculations; however, the findings are applicable to other cavity types, including plasmonic~\cite{hakala2018bose, moilanen2021spatial, rodriguez2013thermalization, wu2016polarons}, photonic crystal~\cite{johnson2001photonic}, microdisk/microring~\cite{zhang2016organic}, defect and gap-based cavities~\cite{urbonas2016zero, scafirimuto2018room, zhu2016quantum, benz2016single, wersall2017observation, baranov2018novel}, among others. The decision to choose a particular material system is motivated by two main reasons: (i) the system must exhibit strong light-matter interaction and exciton-vibration coupling, and even more importantly, it should be capable of supporting polariton Bose--Einstein condensation (BEC); (ii) the chosen molecular system itself should be extensively studied to ensure that our results can be benchmarked against established findings. Organic microcavities based on methyl-substituted conjugated ladder-type polymer (MeLPPP) emerge as the ideal candidates for this analysis. In addition, the temperature-dependent polariton thermalization investigated experimentally~\cite{plumhof2014room} offers invaluable insights for this study shedding light onto the role of low-energy molecular vibrations.


We consider a MeLPPP-based polariton system with the following parameters: cavity dispersion relation $\hbar\omega_{{\rm cav} {\bf k}} = \hbar\omega_{{\bf k}={\bf 0}} + \alpha_{\rm cav} {\bf k}^2$, $\alpha_{\rm cav} = 2.2$~$\rm meV/\mu m^2$; number of molecules within the region of interest $N_{\rm mol}=10^8$ (optically illuminated average area $S=500$~$\rm \mu m^2$)~\cite{zasedatelev2019room, zasedatelev2021single}; the dressed electronic states corresponding to 0-0 absorption peak $\hbar\omega_0 = 2.72~{\rm eV}$ and the inhomogeneous broadening $\hbar\Gamma = 34$~$\rm meV$~\cite{plumhof2014room, xia2023ladder}

Regarding the vibrational degrees of freedom we consider two main low-energy modes associated with the polymer backbone with $\omega_{{\rm v}1} = 48~{\rm cm}^{-1}$, $\omega_{{\rm v}2} = 160~{\rm cm}^{-1}$~\cite{karabunarliev2001franck, snedden2009fluorescence} with corresponding Huang--Rhys factors of $\Lambda_1^2 = 0.7$, $\Lambda_2^2 = 0.5$ respectively~\cite{hildner2009single, baderschneider2016influence}. Although the low-energy modes ($\leq 200$~$\rm cm^{-1}$) present significant experimental challenges for measurement, high-frequency vibrations are readily accessible through conventional Raman spectroscopy. The primary vibrational resonance are as follows: $\omega_{{\rm v}3} = 1320~{\rm cm}^{-1}$, $\omega_{{\rm v}4} = 1568~{\rm cm}^{-1}$, $\omega_{{\rm v}5} = 1604~{\rm cm}^{-1}$~\cite{zasedatelev2021single} and the corresponding Huang-Rhys factors are $\Lambda_3^2 = 0.3$, $\Lambda_4^2 = 0.23$, $\Lambda_5^2 = 0.082$~\cite{guha2003temperature}.
Indeed, in~\cite{guha2003temperature} the Franck--Condon analysis gives the Huang--Rhys factor around $0.65$, whereas we have $\Lambda_3^2+\Lambda_4^2+\Lambda_5^2 = 0.612$.



To extract polariton thermalization rate we proceed with the following steps.
First, we calculate the emission and absorption spectra of the bare MeLPPP layer at different temperatures and extract the net properties of the low-frequency vibrations according to Eq.~(\ref{A1})--(\ref{A2}).
Second, we use the extracted net values of $A_1$ and $A_2$ to estimate polariton thermalization rate Eq.~(\ref{therm k'k}) and (\ref{therm kk'}) as the function of light-matter interaction~(\ref{Rabi}) and the energy of the ground state of polaritons.


Our comparison of the emission and absorption spectra starts by employing three separate approaches: (i) the exact microscopic model as defined by Eq.~(\ref{I_em_1})--(\ref{I_abs_1}); (ii) the reduced model accounting for the net effect of low-energy vibrational modes Eq.~(\ref{I em renormalized})--(\ref{I abs renormalized}); and (iii) the exact model again, but this time excluding all low-energy modes. Figure~\ref{fig: spectra} presents the results of these calculations at two different temperatures, 6~K and 300~K.
Our reduced method agrees well on a quantitative level with the exact microscopic model that includes all mentioned vibrational modes. Furthermore, it is clear that neglecting low-energy modes fails to accurately reproduce the Stokes shift and linewidth in the spectra.

\begin{figure}
\includegraphics[width=0.9\linewidth]{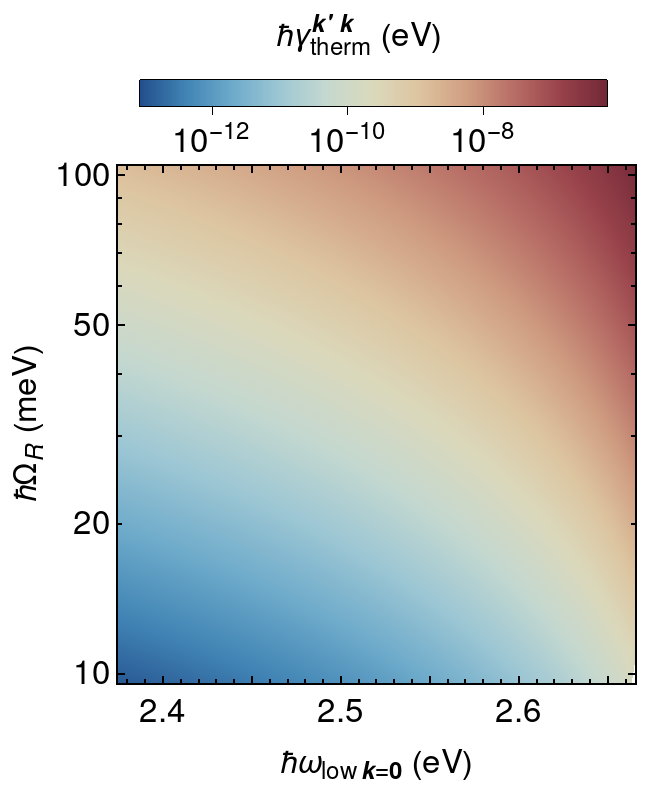}
\caption{
Polariton thermalization rate calculated from the spectral analysis according to Eq.~(\ref{therm k'k}) as the function of light-matter interaction strength (Rabi frequency $\Omega_R$) and the energy of the ground polariton state ($\hbar\omega_{{\rm low}{\bf k}={\bf 0}}$). 
} \label{fig: thermalization_rate}
\end{figure}

The comparison between our theoretical predictions and the experimental absorption and emission spectra reported in the literature demonstrates a quantitative match, including the observed slight asymmetry in the spectral lines~\cite{stampfl1995photoluminescence, pauck1996comparative, hoffmann2010determines}. It is important to note that our theory primarily accounts for the effects of low-energy vibrational (geometric) relaxation and does not take into account factors such as intermolecular interactions or spin multiplicity change. As a result, this theoretical approach is most effective for molecular systems that exhibit a pronounced mirroring of vibronic replicas in their spectra.

The Stokes shift and the linewidth of the 0-0 emission peak can be extracted from Figure~\ref{fig: spectra}(a,b). We have expanded this dataset with additional calculations using the exact model across a range of temperatures from 6~K to 400~K. Figure~(\ref{fig: illustration}) shows the Stokes shift and the linewidth as the function of temperature. Importantly the results obtained from the reduced model using analytical expressions~Eq.~(\ref{omega em renormalized})--(\ref{gamma em renormalized}) are in good agreement with the exact method (Fig.~\ref{fig: illustration}) and 
 experimental data~\cite{hoffmann2010determines}.  Although the Stokes shift remains constant with temperature changes (Fig.~\ref{fig: illustration}a), the linewidth stays constant only at low temperatures and increases linearly with temperature, as illustrated in Fig.~\ref{fig: illustration}b.  Finally, the joint analysis of the emission and absorption spectra at different temperatures allows us to determine the net values $A_1 \approx 18~{\rm meV}$ and $A_2 \approx 200~{\rm meV}^2$ of the low-frequency vibrations, defining polariton thermalization rates Eq.~(\ref{therm k'k})--(\ref{therm kk'}).



\begin{figure}
\includegraphics[width=0.90\linewidth]{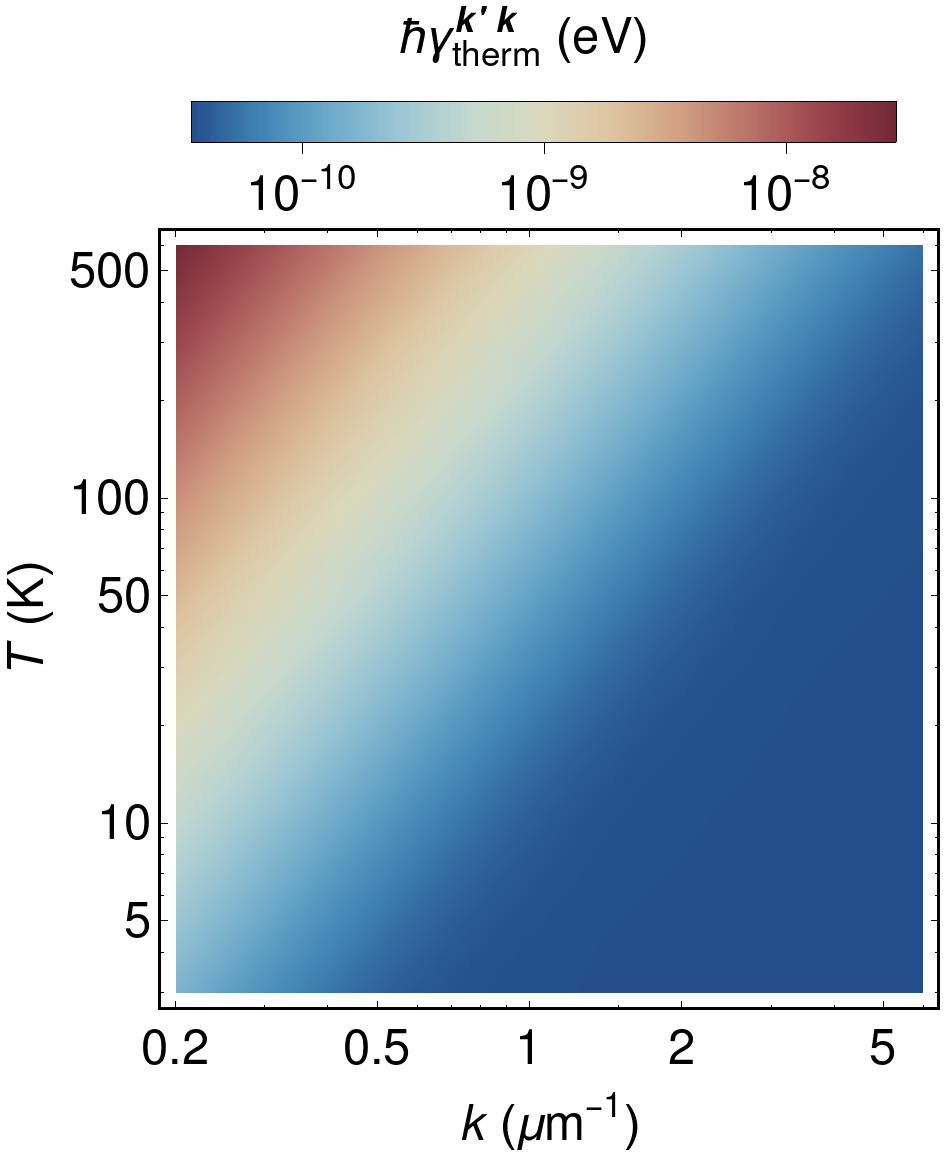}
\caption{
Temperature dependence of the polariton thermalization rate between the ground polariton state (${\bf k}' = {\bf 0}$) and the state with given wavevector $\bf k$ (hor. axis). 
Here we set $\hbar \omega_{{\rm cav}{\bf k}={\bf 0}} = 2.64~{\rm eV}$ and $\hbar \Omega_R = 85~{\rm meV}$.
For the sake of consistency all the other parameters are taken the same as for Fig~\ref{fig: thermalization_rate}.
For these parameters and temperature $T=300~{\rm K}$ Eq.~(\ref{second estimation}) gives the estimation $3.4\cdot 10^{-8}~{\rm eV}$.
} \label{fig: thermalization_rate_temperature}
\end{figure}

Next, we demonstrate how the net values $A_1$ and $A_2$, obtained from the spectral analysis of the bare MeLPPP layer, can be applied to calculate the polariton thermalization rate in practical microcavities~\cite{plumhof2014room,zasedatelev2019room,zasedatelev2021single}. We make use of the equation~(\ref{therm kk'}) to extract thermalization rates between neighboring states in the momentum space, for example $\hbar\bf k' = 0$, and $\hbar\bf k = \hbar\bf k' + \hbar\bf{\delta k}$, with the corresponding energy being $\hbar\omega_{{\rm low}{\bf k}} = \hbar\omega_{{\rm low}{\bf k'}} + 4\pi\alpha_{\rm pol}/S$, where $\alpha_{\rm pol}$ is defined by Eq.~(\ref{dispersion}).
As the cutoff frequency for low-energy vibrations we use ${\omega_M \approx 200~{\rm cm}^{-1} }$.

Figure~\ref{fig: thermalization_rate} presents the extracted values of the polariton thermalization rate as the function of light-matter interaction (Rabi frequency $\Omega_R$) and the energy of the ground polariton state ($\omega_{{\rm low}{\bf k}={\bf 0}}$). 
It should be noted that the polariton thermalization rate has not been directly measured to date and has always been treated as a variable parameter in the fitting of experimental data by microscopic and/or mean-field models~\cite{deng2006quantum, mazza2013microscopic, daskalakis2014nonlinear, hakala2018bose, zasedatelev2021single}.  
Here, we introduce an independent method that provides direct access to the thermalization rate in molecular polariton systems without relying on any adjustable parameters.

Even though thermalization processes reflect the matter behaviour in polariton states, the design of the cavity is expected to have a notable impact, especially regarding aspects tied to the Hopfield coefficients of light-matter states, such as exciton-cavity detuning and mode volume. The general rule of a thumb is that an increase in the material component of the polaritons results in a higher thermalization rate~(Fig.~\ref{fig: thermalization_rate}).
As depicted in Figure~\ref{fig: thermalization_rate} the polariton thermalization rate monotonically increases with the Rabi frequency. Despite the explicit dependence on the number of molecules in Eq.~(\ref{therm rate 1})--(\ref{therm rate 2}) and Eq.~(\ref{therm k'k})--(\ref{therm kk'}), the thermalization rate does not decrease with the increase in $N_{\rm mol}$, because $\Omega_R \propto \sqrt{N_{\rm mol}}$.
Moreover, the thermalization rate grows with the concentration of the molecules due to the increase in material component of the polaritons, which is described by the Hopfield coefficients in Eq.~(\ref{therm rate 1})--(\ref{therm rate 2}).
This observation is in agreement with polariton plasmon systems~\cite{hakala2018bose}.

Temperature is another critical parameter in polariton thermalization. The thermalization rate given by Eq.~(\ref{therm k'k})--(\ref{therm kk'}) exhibits very strong dependence on temperature due to the factor $n_{\rm v}(\Delta \omega_{{\bf kk}'})$. Figure~\ref{fig: thermalization_rate_temperature} demonstrates the peculiar thermalization behavior.
First, the temperature rise leads to an almost linear increase in the thermalization rate to the nearest neighbor polariton state. Secondly, at higher temperatures, the range of polariton states capable of effective thermalization to a specific state expands, or alternatively, it increases the thermalization length, defined as the number of states to which a polariton can efficiently thermalize.
These results explain the bottleneck effect in polariton thermalization observed in organic microcavities at low temperatures~\cite{plumhof2014room}. It underlines the distinct role of thermalization processes in nonequilibrium Bose--Einstein condensation of molecular exciton-polaritons.

\begin{figure}
\includegraphics[width=1\linewidth]{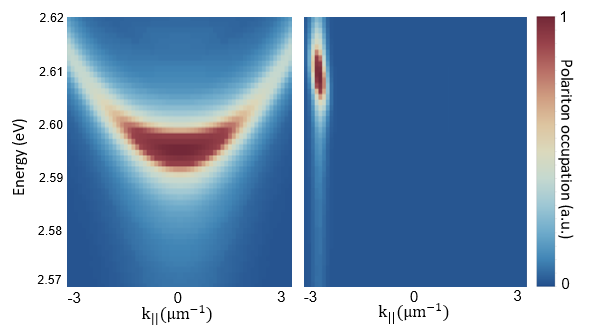}
\caption{
Numerical simulation of the ground (left) and excited state (right) polariton condensation. Here we use microscopic model Ref.~\cite{zasedatelev2021single} 
with experimental parameters from Ref.\cite{baranikov2020all} and effective thermalization rate of $\gamma_{\rm therm} = 5\cdot10^{-10}~{\rm eV}$ extracted from the analytical theory.
} \label{fig: numerics}
\end{figure}

Our theory enables direct access to the polariton thermalization rate from standard spectroscopic measurements of bare molecules. Here we provide joint theoretical analysis for the ground and excited state polariton condensation recently achieved in experiments~\cite{baranikov2020all}. We use numerical model developed in Ref.~\cite{zasedatelev2021single} with the value of thermalization rate obtained here and experimental parameters from Ref.~\cite{baranikov2020all} to simulate $E,k-$distributions of polariton occupation. Figure~\ref{fig: numerics} shows results of numerical simulations for the ground state (left) and excited state $\mathbf{k} = 2.5~\mu m^{-1}$ (right) polariton condensation. It is worth mentioning that the condensation in excited states with high in-plane momenta is quite sensitive to the thermalization rate. Stable condensation in this regime requires a delicate balance between the gain stimulated through the seed beam and the losses mainly coming from finite cavity photon lifetime and thermalization processes favoring downward relaxation to the ground state. Therefore excited state condensation appears to be an excellent testbed to benchmark thermalization theories. The effective polariton thermalization rate extracted from our analytical theory $\gamma_{\rm therm} = 5\cdot10^{-10}~{\rm eV}$ demonstrates quantitative agreement with the experiment~\cite{baranikov2020all} throughout the simulations. Note, the model does not use any free parameters, see Supplementary Information for further details.

\section{Conclusion}

In this work, we investigate the microscopic origins of polariton thermalization in organic systems. We have developed a theoretical framework that encompasses both strong light-matter and exciton-vibration couplings. The main focus of our study is the role of low-energy vibrations in shaping the spectral properties of molecular systems and their interplay with polariton thermalization when strongly coupled to an optical cavity. Analytical expressions for emission and absorption spectra have been derived to effectively incorporate low-frequency vibrational coupling. We introduced net parameters accounting for the ensemble averaged properties of low-energy modes, which can be extracted directly from the Stokes shift ($A_1$) and the temperature-dependent linewidth of the 0-0 vibronic peak in the emission spectrum ($A_2$) of the molecular system without a cavity. In the next step, we established the correspondence between the spectral properties of the molecular system and the polariton thermalization within the cavity.

We have derived a simple analytical expression for the polariton thermalization rate that is proportional to $A_1$ and $A_2$ at high temperatures, meaning a high thermalization rate requires a large Stokes shift and offset linear dependence on the spectral broadening with temperature. Moreover, the polariton thermalization rate depends strongly on temperature and the cavity properties. Finally, we applied the developed formalism to a practical microcavity structure based on the MeLPPP conjugated polymer and calculated the thermalization rate without use of any free parameters. Our results showcase remarkable agreement with recent experimental reports of nonequilibrium polariton condensation in the ground~\cite{plumhof2014room, zasedatelev2019room, zasedatelev2021single} and excited~\cite{baranikov2020all} states, and explain the thermalization bottleneck effect at low temperatures~\cite{plumhof2014room}.

Our research lays the groundwork for understanding nonequilibrium polariton condensation in molecular systems, its interplay with vibrational degrees of freedom, and serves as a guideline for future experimental studies, providing recipes to choose the proper material system and cavity design, as well as conditions to control the thermalization behavior in strongly coupled molecular systems.

\begin{acknowledgments}
A.E.S and Sh.V.Yu. thanks Russian Science Foundation (project No. 20-72-10057).
Sh.V.Yu. thanks the Foundation for the Advancement of Theoretical Physics and Mathematics ``Basis''. A.V.Z. acknowledges support from the
European Union’s Horizon 2020 research and innovation programme under the Marie Sklodowska-Curie grant
agreement No 101030987 (LOREN). 
\end{acknowledgments}

\appendix

\begin{widetext}


\section{Emission spectrum of an organic film} \label{appendix: emission}

The stationary emission spectrum, $I_{\rm em}(\omega)$, can be calculated as follows~\cite{scully1997quantum, carmichael2009open}
\begin{equation} \label{I em def}
I_{\rm em}(\omega) 
= 
C_{\rm em}
\sum_{m=1}^{N_{\rm mol}} 
{\rm Re} \int^{+\infty}_0 e^{-i\omega \tau} \langle \hat \sigma^{(m)\dag}(t+\tau) \hat \sigma^{(m)}(t) \rangle d\tau
\end{equation}
where $t \to \infty$ and $C_{\rm em}$ is the normalization constant.
We set $C_{\rm em}$ such that
\begin{equation} \label{em normalization}
\int_{-\infty}^{+\infty} I_{\rm em}(\omega) d\omega = 1.
\end{equation}

We assume that the weak incoherent pumping supplies the energy to the dressed excitons and do not affect the dressed molecular vibrations. 
In this case, we can separate contribution of these two subsystems in the two-time correlator form Eq.~(\ref{I em def})
\begin{equation}  \label{factor_sigma}
\langle \hat \sigma^{(m)\dag}(t+\tau) \hat \sigma^{(m)}(t) \rangle 
\approx 
\langle \hat S^{(m)\dag} (t+\tau)\hat S^{(m)}(t)\rangle \prod_{j=1}^{N_{\rm vib}}  \langle \hat D_{j}^{(m)\dag} (t+\tau)\hat D_{j}^{(m)}(t)\rangle,
\end{equation}

The analysis of the dynamics of the excitons at $t \to +\infty$ gives~\cite{reitz2019langevin}
\begin{equation} \label{StSt+tau}
\langle \hat S^\dag (t+\tau)\hat S(t)\rangle=pe^{-(\gamma_{\rm diss}/2-i\omega_0)\tau},
\end{equation}.

To obtain $\langle \hat D^{(m)\dag}_{j}(t+\tau) \hat D^{(m)}_{j}(t) \rangle$ at $t\to +\infty$, we consider the correlator 
$${\langle \hat R^{(m)}_{j}(t+\tau, \mu_1, \mu_2) \hat D^{(m)}_{j}(t) \rangle},$$ where $\hat R^{(m)}_{j}(t+\tau, \mu_1, \mu_2)=e^{\mu_1 \hat B^{(m)\dag}_j(t+\tau)}e^{-\mu_2 \hat B^{(m)}_j(t+\tau)}$. 
Using the Lindblad equation~(\ref{MasterEquation}) and the quantum regression theorem~\cite{scully1997quantum, carmichael2009open}, we obtain an equation for this correlator
\begin{multline} \label{diff_eq}
\frac{\partial}{\partial \tau}\langle \hat R^{(m)}_{j}(t+\tau, \mu_1, \mu_2)\hat D^{(m)}_{j}(t)\rangle
=
\\
\left(-(\gamma_{{\rm v}j}-i \omega_{{\rm v}j})\mu_1\frac{\partial}{\partial \mu_1}-(\gamma_{{\rm v}j}+i \omega_{{\rm v}j})\mu_2\frac{\partial}{\partial \mu_2}-2\gamma_{{\rm v}j}n_{{\rm v}j}\mu_1\mu_2\right)\langle \hat R^{(m)}_{j}(t+\tau, \mu_1, \mu_2)\hat D^{(m)}_{j}(t)\rangle
.
\end{multline}
with the initial conditions
\begin{equation}
\langle \hat R^{(m)}_{j}(t, \mu_1, \mu_2) \hat D^{(m)}_{j}(t) \rangle
=
\langle e^{\mu_1 \hat B^{(m)\dag}_j(t)}e^{-\mu_2 \hat B^{(m)}_j(t)} e^{-\Lambda_j (\hat B_j^{(m)\dag}(t) - \hat B_j^{(m)})(t)}\rangle
\end{equation}
One can show that
\begin{equation}
\langle \hat R^{(m)}_{j}(t, \mu_1, \mu_2) \hat D^{(m)}_{j}(t) \rangle
=
e^{-\Lambda_j^2/2}e^{\mu_2\Lambda_j}e^{-n_{{\rm v}j}(\Lambda_j-\mu_1)(\Lambda_j-\mu_2)}
.
\end{equation}
From here, we can find the solution to the equation~(\ref{diff_eq})
\begin{equation}
\langle \hat R^{(m)}_{j}(t+\tau, \mu_1, \mu_2) \hat D^{(m)}_{j}(t) \rangle
=
e^{-\Lambda_j^2/2}e^{-n_{{\rm v}j}\Lambda_j^2}e^{-n_{{\rm v}j}\mu_1\mu_2}e^{n_{{\rm v}j}\Lambda_j\mu_1 e^{-(\gamma_{{\rm v}j}-i \omega_{{\rm v}j})\tau}}e^{(1+n_{{\rm v}j})\Lambda_j\mu_2 e^{-(\gamma_{{\rm v}j}+i \omega_{{\rm v}j})\tau}}.
\end{equation}
Using equality $\langle \hat D^{(m)\dag}_{j}(t+\tau) \hat D^{(m)}_{j}(t) \rangle = e^{-\Lambda_j^2/2}\langle \hat R^{(m)}_{j}(t+\tau, \Lambda_j, \Lambda_j) \hat D^{(m)}_{j}(t) \rangle$, we finally obtain
\begin{equation} \label{DtDt+tau}
\langle \hat D^{(m)\dag}_{j}(t+\tau) \hat D^{(m)}_{j}(t) \rangle 
=
e^{-n_{{\rm v}j}\Lambda_j^2}e^{n_{{\rm v}j}\Lambda_j^2 e^{-(\gamma_{{\rm v}j}-i \omega_{{\rm v}j})\tau}}
e^{-(1+n_{{\rm v}j})\Lambda_j^2}e^{(1+n_{{\rm v}j})\Lambda_j^2 e^{-(\gamma_{{\rm v}j}+i \omega_{{\rm v}j})\tau}}.
\end{equation}

Substitution of Eq.~(\ref{StSt+tau}) and Eq.~(\ref{DtDt+tau}) into Eq.~(\ref{factor_sigma}) allows us to transform Eq.~(\ref{I em def}) into
\begin{multline} \label{I_em}
I_{\rm em}(\omega) 
= 
\sum_{k'_1,...,k'_{N_{\rm vib}}} 
z_{k'_1}(n_{{\rm v}1}\Lambda_1^2)
...
z_{k'_{N_{\rm vib}}}(n_{{\rm v}N_{\rm vib}}\Lambda_{N_{\rm vib}}^2)
\\
\sum_{k_1,...,k_{N_{\rm vib}}}  
z_{k_1}((1+n_{{\rm v}1})\Lambda_1^2)
...
z_{k_M}((1+n_{{\rm v}N_{\rm vib}})\Lambda_{N_{\rm vib}}^2)
\\
\frac{1 }{ N_{\rm mol}}
\sum_{m=1}^{N_{\rm mol}}
\frac{
\gamma_{\rm diss}/2 + \gamma_{{\rm v}\Sigma}  
}{ 
(\omega - \omega_0^{(m)} + \omega_{{\rm v}\Sigma})^2 + (\gamma_{\rm diss}/2 + \gamma_{{\rm v}\Sigma})^2
},
\end{multline}
where we introduced
\begin{equation}
\omega_{{\rm v}\Sigma}
=
\sum_{j=1}^{N_{\rm vib}}\omega_{{\rm v}j}(k_j-k'_j),
\end{equation}
\begin{equation}
\gamma_{{\rm v}\Sigma}
=
\sum_{j=1}^{N_{\rm vib}}\gamma_{{\rm v}j}(k_j+k'_j).
\end{equation}

Eq.~(\ref{I_em}) at $T=0$ ($n_{{\rm v}j}=0$) coincides with one obtained in~\cite{reitz2019langevin} via Heisenberg--Langevin approach.

To obtain Eq.~(\ref{I_em_1}), we use the arguments from Section~\ref{subsec: em and abs spectra}.
Namely, due to high molecular density of organic thin films, the distribution of the energies of the excitons can be treated as continuous.
We assume that this distribution is normal with the standard deviation $\Gamma$.
We also assume that $\Gamma \gg \gamma_{\rm diss}/2 + \gamma_{{\rm v}\Sigma}$.
In this case, the inhomogeneous broadening leads to the Gauss lineshapes of the spectral peaks~\cite{knapp1984lineshapes, spano2009determining, guha2003temperature, borrelli2014theoretical, ostroverkhova2016organic}.
Thus, from Eq.~(\ref{I_em}), we obtain Eq.~(\ref{I_em_1}).

To obtain Eq.~(\ref{I em renormalized}) from Eq.~(\ref{I_em_1}), we change the sums in Eq.~(\ref{I_em_1}) corresponding to the low-frequency vibrations as follows
\begin{equation} \label{sum to int}
\sum_{n=0}^{+\infty} 
\frac{e^{-x} x^n }{ n!} 
... 
=
\left.
\frac{2 }{ 1 + {\rm efr}\left(\sqrt{x/2}\right)}
\frac{1 }{ \sqrt{2 \pi x}}
\int_{0}^{+\infty}dy e^{-(y-x)^2/2x}
...
\right|_{n \to y}
\end{equation}

\section{Absorption spectrum of an organic film}  \label{appendix: absorption}

The absorption spectrum, $I_{\rm abs}(\omega)$, can be calculated from~\cite{loudon2000quantum}
\begin{equation} 
I_{\rm abs}(\omega_\Omega) 
= 
-
C_{\rm abs}
\sum_{m=1}^{N_{\rm mol}}
{\rm Im} \langle \hat \sigma^{(m)}(t)e^{i\omega_\Omega t} \rangle/\Omega
,
\end{equation}
where ${t \to +\infty}$ and $C_{\rm abs}$ is the normalization constant.
We set $C_{\rm abs}$ such that
\begin{equation} \label{abs normalization}
\int_{-\infty}^{+\infty} I_{\rm abs}(\omega) d\omega = 1. 
\end{equation}

In the basis of dressed states, the expression for the absorption spectrum has the form
\begin{equation}
I_{\rm abs}(\omega_\Omega)
=
-
\frac{ C_{\rm abs} }{ \Omega }
\sum_{m=1}^{N_{\rm mol}}{\rm Im} 
\left\langle 
\prod_{j=1}^{N_{\rm vib}} \hat D_j^{(m)}(t) \hat S^{(m)}(t)e^{i\omega_\Omega t} 
\right\rangle  
\end{equation}

The Lindblad equation~(\ref{MasterEquation}) allows us to write the Heisenberg-Langevin equation for the operator $\hat S(t)$
\begin{equation} \label{S equation}
\frac{d}{dt}\hat S^{(m)}(t)=-(\gamma_{\rm diss}/2+i\omega_0^{(m)})\hat S^{(m)}(t)-i\Omega\hat D^{(m)\dag}(t)e^{-i \omega_\Omega t}+\hat F^{(m)}(t),
\end{equation}
where $\hat F^{(m)}(t)$ is the noise acting on the dressed electronic state due to the interaction of the molecule with the environment, while $\langle \hat F^{(m)}(t) \rangle = 0$.
In the equation~(\ref{S equation}), we discarded the term that is nonlinear in $\hat S^{(m)}(t)$, assuming that the external field is weak.

We find the operator $\hat S(t)$ by integrating the equation~(\ref{S equation}), then we substitute the result into the expression for $I_{\rm abs}(\omega_\Omega)$, and obtain
\begin{equation} \label{appendix: I_abs}
I_{\rm abs}(\omega_\Omega)
=
C_{\rm abs}
\sum_{m=1}^{N_{\rm mol}}
{\rm Re}  
\int_0^{+\infty}
\prod_{j=1}^{N_{\rm vib}}
\left\langle 
\hat D_j^{(m)}(t)\hat D_j^{(m)\dag}(t')
\right\rangle 
e^{i \omega_{\Omega}(t-t')}e^{-(\gamma_{\rm diss}/2+i\omega_0)(t-t')}dt'.  
\end{equation}
Here, we assume that the noise acting on the dressed electronic state is uncorrelated with the dressed vibrational states of the molecule, i.e. $\langle \hat D_j^{(m)}(t) \hat F^{(m)}(t) \rangle = \langle \hat D_j^{(m)}(t) \rangle \langle \hat F^{(m)}(t) \rangle = 0$.

The correlator ${\langle \hat D_j^{(m)}(t)\hat D_j^{(m)\dag}(t')\rangle}$ can be find similarly to how we found $\langle \hat D_j^{( m)\dag}(t+\tau)\hat D_j^{(m)}(t)\rangle$ in Section~\ref{appendix: emission}.
Thus, we obtain
\begin{multline} \label{I_abs}
I_{\rm abs}(\omega) 
= 
\sum_{k'_1,...,k'_{N_{\rm vib}}} 
z_{k'_1}(n_{{\rm v}1}\Lambda_1^2)
...
z_{k'_{N_{\rm vib}}}(n_{{\rm v}N_{\rm vib}}\Lambda_{N_{\rm vib}}^2)
\\
\sum_{k_1,...,k_{N_{\rm vib}}}  
z_{k_1}((1+n_{{\rm v}1})\Lambda_1^2)
...
z_{k_M}((1+n_{{\rm v}N_{\rm vib}})\Lambda_{N_{\rm vib}}^2)
\\
\frac{1 }{ N_{\rm mol}}
\sum_{m=1}^{N_{\rm mol}}
\frac{
\gamma_{\rm diss}/2 + \gamma_{{\rm v}\Sigma}  
}{ 
(\omega - \omega_0^{(m)} - \omega_{{\rm v}\Sigma})^2 + (\gamma_{\rm diss}/2 + \gamma_{{\rm v}\Sigma})^2
}.
\end{multline}
At zero temperature ($n_{{\rm v}j}=0$) Eq.~(\ref{I_abs}) coincides with the one obtained in~\cite{reitz2019langevin}.

As we did in the Section~\ref{appendix: emission} we can take into account inhomogeneous broadening and obtain~Eq.~(\ref{I_abs_1}) from Eq.~(\ref{I_abs}).
Using Eq.~(\ref{sum to int}), we can effectively incorporate low-frequency vibrational modes and obtain Eq.~(\ref{I em renormalized}) from Eq.~(\ref{I_abs_1}).

\section{Numerical simulation of Bose--Einstein condensation} \label{appendix: numerical}

Simulations were carried out within the framework of Lindblad master equation~\cite{zasedatelev2021single}:
\begin{equation}
    \frac{d\hat{\rho}}{dt}=\frac{i}{\hbar}\left[\hat{\rho},\hat{H}\right]+\hat{L}_{\rm up}(\hat{\rho})+\hat{L}_{\rm low}(\hat{\rho})+\hat{L}_{\rm vib}(\hat{\rho})+\hat{L}_{\rm therm}(\hat{\rho})+\hat{L}_{\rm pump}(\hat{\rho})+\hat{L}_{\rm seed}(\hat{\rho})
\end{equation}

Tracing out polariton occupation we can adiabatically exclude vibrational degrees of freedom and obtain the following discrete set of rate equations within the region of interest  \(|k|<3~{\rm\mu m}^{-1}\).

\begin{equation}
    \begin{gathered}
        \frac{dn_P}{dt}=-\gamma_P n_P+\kappa_P(t)-\sum_j \Gamma_j^P n_P (n_j+D_j)\\
        \frac{dn_i}{dt}=-\gamma_i n_i+\kappa_{\rm seed}(k_i,t)+ \Gamma_i^P n_P (n_i+D_i) +\\ +\sum_j\left[ \gamma_{\rm therm}^{j\rightarrow i} n_j (n_i+D_i)-\gamma_{\rm therm}^{i\rightarrow j} n_i (n_j+D_j)\right]
    \end{gathered}
\end{equation}

To access spectral properties we apply quantum regression theorem~\cite{zasedatelev2021single} to generate equations for two-time correlator functions in a form:
\(\langle \hat{s}_{\boldsymbol{k}}^\dagger(t+\tau)\hat{s}_{\boldsymbol{k}}(t) \rangle\), where amplitudes of polariton states within region of interest defined as $\hat{s}_{\boldsymbol{k}} = \langle \hat{s}_{\boldsymbol{k}} \rangle$ ($\textit{mean-field approximation}$). On the next step, we solve them numerically and apply Fourier transformation to the two-time correlators at each mode followed by integration over time.

We numerically solve the differential equations for \(N=31\) modes at the lower polariton branch including the ground state. The parameters for the model adopted from experimental data Ref.~\cite{baranikov2020all}, the main ones are the following:

\begin{itemize}
    \item Light-matter interaction: \(\omega_{\rm exc}=2.72~{\rm eV}\), \(\omega_{\rm cav}=2.64~{\rm eV}\), \(\Omega_{R}=85~{\rm meV}\), \(\alpha_{\rm cav}=2.2~{\rm meV}\cdot{\rm \mu m^2}\)
    \item Decay rates: \(\gamma_{\rm cav}=4.4~{\rm meV}\), \(\gamma_{\rm exc}=60~{\rm meV}\)
    \item High-energy vibrations: \(\omega_{\rm vib}=0.199~{\rm eV}\), \(\gamma_{\rm vib}=2.5~{\rm meV}\), \(g=0.5~{\rm meV}\)
    \item Optical pumping: \(\omega_{P}=2.8~{\rm eV}\), \(\tau_{P}^{FWHM}=200~{\rm fs}\), \(P=2P_{\rm th}\)
    \item Polariton seed: \(k_{\rm seed}=2.55~{\rm \mu m^{-1}}\), \(\sigma_{\rm seed}=0.2~{\rm \mu m^{-1}}\), \(\tau_{\rm seed}^{\rm FWHM}=200~{\rm fs}\)
\end{itemize}

The estimation~Eq.~(\ref{therm k'k}) gives the thermalization rate for the neighbouring states with the wave vectors $\bf k$ and $\bf k'$
\begin{equation} \label{estimation for the neighbouring states}
\left(
\gamma_{\rm therm}^{{\bf k'}{\bf k}}
\right)_{k_BT\gtrsim \hbar\omega_M}^{\rm est} \simeq 1.4\cdot 10^{-8}~({\rm eV}).   
\end{equation}
where we set $A_1 \approx 18~{\rm meV}$, $A_2 \approx 200~{\rm meV}^2$ and $\omega_M = 200~{\rm cm}^{-1}$~(see main text Section~\ref{sec: example}).
The numerical simulations ignore some of the details in the thermalization process, accounting for them only effectively. For example, here we assume that thermalization rate is constant within the region of interest, i.e. it does not depend on the in-plane momentum $\hbar k$: \(\gamma_{\rm therm}^{\boldsymbol{k}_1 \rightarrow \boldsymbol{k}_2}=\gamma_{\rm therm}\) when \(\omega_1> \omega_2\) and is equal to \(\gamma_{\rm therm}\exp{(-\hbar (\omega_2-\omega_1)/k_B T)}\) otherwise.
Therefore, we cannot use our estimation~Eq.~(\ref{estimation for the neighbouring states}) directly.
Nevertheless, by averaging over many thermalization steps of Eq.~(\ref{estimation for the neighbouring states}), our analytical theory can provide the effective thermalization rate per step of~$\gamma_{\rm therm}\simeq 5\cdot10^{-10}~{\rm eV}$ that we use in the numerical simulations.

\end{widetext}

\bibliography{main}

\end{document}